\documentclass[amsmath,amssymb,aps,preprintnumbers,prd,twocolumn,superscriptaddress,showpacs]{revtex4}

\usepackage{multirow}
\usepackage[dvips]{graphicx}
\usepackage{times}
\usepackage{color}
\usepackage[dvipdfm, colorlinks=true, pdfstartview=FitV, linkcolor=red, citecolor=blue, urlcolor=blue]{hyperref}
\usepackage[normalem]{ulem} 

\newcommand{\comment}[1]{}
\newcommand{\nn}{\nonumber}
\newcommand{\sla}[3]{\hspace{#1mm}\not\hspace{#2mm}{#3}\hspace{-#2mm}}

\newcommand{\abs}[1]{|\mathbf{#1}|}

\newcommand{\sub}[1]{{\scriptscriptstyle \mathrm{#1}}}

\renewcommand\sout{\bgroup \color{blue} \ULdepth=-.5ex \ULset}
\begin{document}
\preprint{RIKEN-QHP-50, KUNS-2452}

\title{
Neutrino spectral density at electroweak-scale temperature}

\author{Kohtaroh Miura}
\email[]{miura@kmi.nagoya-u.ac.jp}
\affiliation{INFN-Laboratori Nazionali di Frascati, I-00044, Frascati(RM), Italy}
\author{Yoshimasa Hidaka}
\email[]{hidaka@riken.jp}
\affiliation{Theoretical Research Division, Nishina Center, RIKEN, Wako 351-0198, Japan}
\author{Daisuke Satow}
\email[]{daisuke.sato@riken.jp}
\affiliation{Theoretical Research Division, Nishina Center, RIKEN, Wako 351-0198, Japan}
\affiliation{Department of Physics, Kyoto University, Kyoto 606-8502, Japan}

\author{Teiji Kunihiro}
\email[]{kunihiro@ruby.scphys.kyoto-u.ac.jp}
\affiliation{Department of Physics, Kyoto University, Kyoto 606-8502, Japan}
\date{\today}
\begin{abstract}
Motivated by the scenario of resonant leptogenesis in which
lepton number creation in the electroweak-scale is relevant,
we investigate the spectral properties and possible collective nature
of the standard model neutrinos
at electroweak-scale temperature ($T$).
We adopt the $R_{\xi}$ gauge fixing,
which includes the unitary gauge as a limiting case,
and allows us to study the broken as well as the restored phases of
the gauge symmetry in a unified way.
We show that the spectral density of the neutrino has
a three-peak structure in the low-momentum region due to the 
scattering with the thermally excited particles
(i.e., Landau damping) for $T\gtrsim M_{\sub{W,Z}}(T)$ with
$M_{\sub{W,Z}}(T)$ being the weak-boson masses in the plasma.
The three peaks are identified with a novel ultrasoft mode,
the usual quasiparticle, and antiplasmino modes.
Varying the gauge-fixing parameter $\xi$,
we show that the three-peak structure appears independently 
of the gauge fixing and thus has a physical significance.
We discuss possible implications of
the neutrino spectral density obtained in the present work 
on particle cosmology, in particular in the
context of resonant leptogenesis.
\end{abstract}
\maketitle
\section{Introduction}\label{sec:intro}
Thermodynamic properties of gauge theories are of common interest
in various fields of physics ~\cite{LeBellac,Kapusta},
including the baryogenesis and leptogenesis,
the physics of quark-gluon plasma (QGP),
neutrinos physics in big-bang nucleosynthesis
and in compact stars, and many others.
A novel ingredient in the dynamics at finite temperature ($T$) 
is due to the existence of thermally excited particles, 
with which a probe particle is scattered in the classical level,
leading to a damping known as Landau damping and a modification
of the spectral properties of the probe particle.
In this paper, partly motivated by the recent development of 
the resonant leptogenesis scenario \cite{Pilaftsis,Pilaftsis:2005rv},
we study the spectral properties of neutrinos at finite $T$,
specifically in the electroweak-scale, in a way where
the gauge (in)dependence is manifestly checked.

Although the standard model (SM) seems to fail to explain
the observed baryon asymmetry of the universe (BAU),
an extension of the SM by adding right-handed Majorana neutrinos
($N_{\sub{R}}$) may have a chance to account for the BAU.
Indeed, a thermal leptogenesis scenario
based on such an extended model
\cite{Fukugita:1986hr,LGene_Review}
gives a simple and promising account of the BAU.
In this scenario,
a decay of $N_{\sub{R}}$ out of equilibrium
via Yukawa interactions
generates a net lepton number,
which is partially converted into the baryon numbers
through sphaleron processes
\cite{Kuzmin:1985mm,Klinkhamer:1984di,Arnold:1987mh}
in the electroweak phase transition.
Furthermore, if the mass difference between
two right-handed neutrinos
is on the order of their $CP$-violating decay width,
the $CP$ asymmetry is 
dynamically enhanced and so is leptogenesis
\cite{Pilaftsis,Pilaftsis:2005rv}.
This resonant leptogenesis scenario implies that 
$N_{\sub{R}}$ with a TeV-scale or electroweak-scale mass
is relevant to the BAU and thus
has opened a possibility to create
a lepton number at electroweak-scale temperature.

\begin{figure}[ht]
\begin{center}
\includegraphics[width=3.0cm]{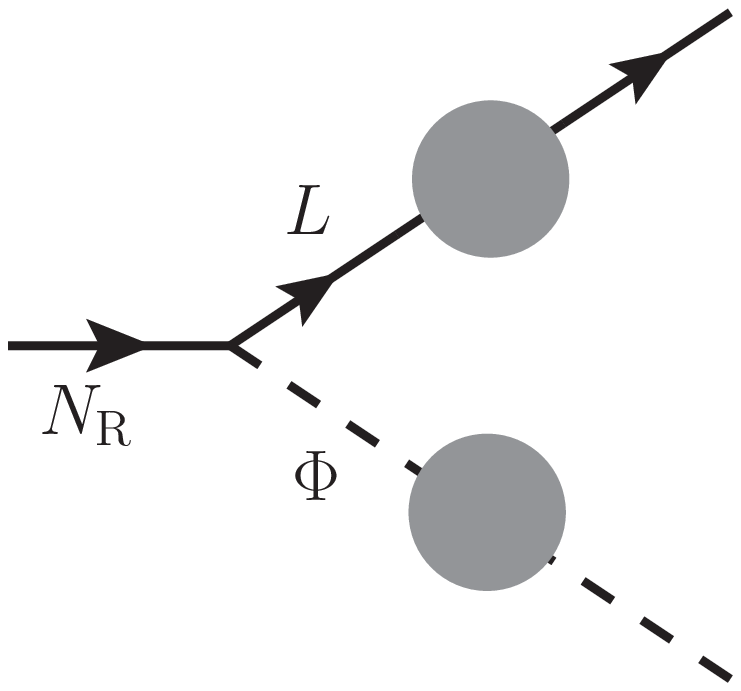}
\caption{A decay process of
a right-handed Majorana neutrino ($N_{\mathrm{R}}$)}
to a standard model lepton $L=(\nu,l_{\sub{L}})^{T}$ and
a standard model Higgs doublet $\Phi$.
\label{Fig:NR_decay}
\end{center}
\end{figure}

Figure~\ref{Fig:NR_decay} shows 
the $CP$-violating decay process of $N_{\sub{R}}$,
which gives one of the most important contributions
to lepton number creation, where
both the Higgs particle and leptons are involved.
In the present work,
we suppose that the relevant processes occur
at finite $T$ of the electroweak-scale.
In this $T$ region,
the following two finite-$T$ effects come into play:
1)~the Landau damping of the Higgs particle and leptons
by thermally excited particles including massive gauge bosons;
2)~the $T$ dependence of the masses of
the involved particles, 
generated
by the Englert-Brout-Higgs-Guralnik-Hagen-Kibble mechanism
\cite{EBHGHK},
which will be called simply
the Higgs mechanism from now on in this paper.
If the standard model leptons have nontrivial
spectral properties due to the aforementioned two effects,
lepton number creation
via the $N_{\sub{R}}$ decay in Fig.~\ref{Fig:NR_decay}
will be significantly modified.

The important role of the Landau damping for
modifying the fermion spectral properties 
is known and established
in QED and QCD in the context of QGP physics.
A fermion interacting with thermally excited gauge bosons and antifermion
admits a novel collective excitation mode
known as a (an anti) plasmino
as well as the normal quasiparticle excitation with thermal mass
\cite{quasi_particle}.
Such collective excitations in the QGP at extremely high $T$ 
are shown to exist and their dispersion relations are obtained in a gauge-invariant way,
at least at one-loop order,
which is now known as the hard thermal loop (HTL)
approximation \cite{HTL}. 
A full spectral property beyond the dispersion relations 
of massive electrons in QED plasma
is investigated in Ref.~\cite{Baym:1992eu},
where a physical account of the interplay
between the electron-mass and temperature-effects
that leads to a Landau damping were presented.

As first discussed in Ref.~\cite{DOlivo:1992vm},
we expect that
the leptons appearing in Fig.~\ref{Fig:NR_decay}
should also admit nontrivial collective excitations at finite $T$.
Recently, the neutrino spectral properties
at electroweak-scale temperature
in the broken phase have been investigated
by using the HTL approximation in the unitary gauge
\cite{Boyanovsky:2005hk}.
A remarkable finding there was the possible existence of
a novel branch of the dispersion relation with
a collective nature in the ultrasoft-energy region.
Here, an important question is whether
such an ultrasoft mode is physically significant with regards to
the gauge invariance, which cannot be checked within the unitary gauge.
In fact, the gauge (in)dependence of the spectral properties of neutrinos
in the ultrasoft-energy region
has not been fully investigated yet,
except for those
 at $T=1$--$10$ MeV~\cite{DOlivo:1992vm}.

Here, we note that the gauge (in)dependence of
the spectral properties of massless fermions (quarks)
in the  ultrasoft-energy region
has been discussed in the physics of QGP.
It was shown \cite{3peakQGP,Kitazawa:2007ep} that the spectral function of 
massless quarks coupled to 
massive vector soft modes in QGP
can have  a novel peak in the  ultrasoft-energy region
as well as those corresponding to the normal and the antiplasmino modes,
and it was later demonstrated~\cite{Satow:2010ia} that 
their properties are gauge independent
by using the Stueckelberg formalism~\cite{Stueckel_O,Stueckel_R}.

Now, one should recognize that 
the relevant ingredients for the description
of hot neutrinos at the electroweak-scale temperature
are the same as those in Ref.~\cite{Satow:2010ia},
{\em i.e.}, a massless fermion and a massive vector boson with
a mass comparable with $T$.
Thus, a natural and intriguing question is whether 
the neutrino spectral density at $T$ of the electroweak-scale
should also develop a three-peak or three-bump structure
including an ultrasoft mode without gauge dependence,
as quarks do in the QGP
\cite{3peakQGP,Kitazawa:2007ep,Satow:2010ia,HaraNemo:2008,Qin:2010pc,Nakkagawa:2012}.

In the present work,  we investigate
in a gauge-invariant way the spectral properties 
of neutrinos at weak coupling, 
without restricting ourselves to their dispersion relations, 
in the full energy-momentum plane
at finite $T$ around the electroweak-scale.
We show that the neutrino develops
an ultrasoft excitation mode as well as
the normal and plasmino ones
without recourse to the HTL approximation.
We shall employ the $R_{\xi}$ gauge~\cite{Fujikawa:1972fe}
for examining the possible gauge-fixing dependence
of the spectral density in both broken and symmetric phases.
Furthermore, we also discuss
the implications of the resulting neutrino spectral properties
to the cosmology by considering
the electroweak-scale resonant leptogenesis.

This paper is organized as follows.
In the next section,
we introduce the Lagrangian of the electroweak theory 
in the $R_{\xi}$ gauge~\cite{Fujikawa:1972fe} and 
the basic quantities  to
evaluate the spectral properties of neutrinos
in field theory at finite temperature.
In Sec.~\ref{sec:Res},
we  show the numerical results for the self-energy
and the spectral density
of neutrinos in the broken phase.
The results will be
compared with the unitary-gauge HTL results
\cite{Boyanovsky:2005hk}.
In Sec.~\ref{sec:SDens_M0},
we  extend our study into the symmetric phase,
and investigate the fate of collective modes for neutrinos.
Then in Sec.~\ref{sec:discuss},
we discuss possible implications of
the neutrino spectral density obtained in the present work 
in the context of resonant leptogenesis.
Finally, we make concluding remarks in Sec.~\ref{sec:sum}.
Appendices \ref{app:EW_params}--\ref{App:SEne_M0}
are devoted to explaining technical details.

\section{Preliminaries}\label{sec:EWT_SDens}

\subsection{Electroweak theory at finite temperature in $R_{\xi}$ gauge}
\label{subsec:EWPT_Rxi}
The Lagrangian to be considered is composed of
the lepton $L$,
the Higgs $\Phi$,
and the gauge boson
$\mathcal{G}_{\mu}=\mathcal{A}_{\mu},\mathcal{B}_{\mu}$ sectors,
\begin{align}
&\mathcal{L}[\Phi,L,\mathcal{G}_{\mu}]
=
\mathcal{L}_{\sub{L}}[L,\mathcal{G}_{\mu}]
+ \mathcal{L}_{\sub{H}}[\Phi,\mathcal{G}_{\mu}]
+ \mathcal{L}_{\sub{G}}[\mathcal{G}_{\mu}]
\ ,\label{eq:L_tot}
\end{align}
up to gauge fixing and
Faddeev-Popov ghost terms.
The Higgs sector is given by
\begin{align}
\mathcal{L}_{\sub{H}}
&=\bigl(\mathcal{D}_{\mu}\Phi\bigr)^{\dagger}
\bigl(\mathcal{D}^{\mu}\Phi\bigr)
+ \mu^2_0\Phi^{\dagger}\Phi
- \lambda\bigl(\Phi^{\dagger}\Phi\bigr)^2\ ,\label{eq:LH}\\
\mathcal{D}_{\mu}\Phi
&=      \biggl(
		\partial_{\mu}\mathbf{1}
		- ig\frac{\sigma^a\mathcal{A}_{\mu}^a}{2}
		- \frac{ig^{\prime}}{2}\mathcal{B}_{\mu}\mathbf{1}
	\biggr)\Phi \ ,
\end{align}
where $\sigma^{a=1,2,3}$ are Pauli matrices,
and $\mathcal{A}_{\mu}$ and
$\mathcal{B}_{\mu}$ are the gauge fields
of $\mathrm{SU}_\sub{W}(2)$ and $\mathrm{U}_\sub{Y}(1)$ groups,
respectively.
The field $\Phi$ represents the weak doublet scalar,
which can be parametrized as
\begin{align}
\Phi(x)
=	\frac{1}{\sqrt{2}}
\begin{bmatrix}
\phi^{2}(x)+i\phi^{1}(x)\\
v+h(x)-i\phi^3(x)
\end{bmatrix}
\ .
\end{align}
Here, the constant $v$
represents the vacuum expectation value (VEV).
The fluctuations  
of the Higgs field and the Nambu-Goldstone modes
are denoted by $h(x)$ and $\phi^a(x)$, respectively,
the latter of which is to be eaten by the gauge fields.

The gauge-fixing term in the $R_{\xi}$ gauge reads
\begin{align}
\mathcal{L}_{\sub{GF}}
&=
-\frac{1}{2\xi}
\Bigl(
\bigl(F^a_{\sub{A}}(x)\bigr)^2
+ \bigl(F_{\sub{B}}(x)\bigr)^2
\Bigr)\ ,\label{eq:L_G_GF}
\end{align}
where
\begin{align}
&F^a_{\sub{A}}(x)
= \partial_{\mu}\mathcal{A}^{\mu,a}(x)
+ \xi m_{\sub{A}} \phi^a(x)\ ,\label{eq:GFA}\\
&F_{\sub{B}}(x)
= \partial_{\mu}\mathcal{B}^{\mu}(x)
- \xi m_{\sub{B}}\phi^3(x)\ ,\label{eq:GFB}
\end{align}
with
$(m_{\sub{A}},m_{\sub{B}})
=(g\mu_0/(2\sqrt{\lambda}),g^{\prime}\mu_0/(2\sqrt{\lambda}))$.
The gauge-fixing conditions
are specified by the parameter $\xi$.

From the gauge-boson sector $\mathcal{L}_{\sub{G}}$
and the gauge-fixing term $\mathcal{L}_{\sub{GF}}$,
the bare propagators of the gauge bosons
are obtained as a sum of the unitary gauge
and the $\xi$-dependent parts,
\begin{align}
G_{\mu\nu}(q)
&=
G_{\mu\nu}^{\sub{U}}(q) + G_{\mu\nu}^{\xi}(q)
\ ,\label{eq:G}\\
G_{\mu\nu}^{\sub{U}}(q)
&=
-\sum_{t=\pm}
\frac{t\bigl(g_{\mu\nu}-q_{\mu}q_{\nu}/M^2(T)\bigr)}
{2E_{\mathbf{q}}(M)\bigl(q_0-tE_{\mathbf{q}}(M)\bigr)}
\ ,\label{eq:GU}\\
G_{\mu\nu}^{\xi}(q)
&=
\sum_{t=\pm}
\frac{t\cdot q_{\mu}q_{\nu}/M^2(T)}
{2E_{\mathbf{q}}^{\xi}(M)
\bigl(q_0-tE_{\mathbf{q}}^{\xi}(M)\bigr)}
\ ,\label{eq:Gxi}
\end{align}
where
\begin{align}
E_{\mathbf{q}}(M)\equiv \sqrt{\abs{q}^2+M^2(T)}\ ,
E_{\mathbf{q}}^{\xi}(M)\equiv E_{\mathbf{q}}(\sqrt{\xi}M)
\ ,\label{eq:E}
\end{align}
and $M(T)$ represents
the $W$ boson, $Z$ boson or photon mass
at finite $T$:
$M(T)=M_{\sub{W}}(T),~M_{\sub{Z}}(T),~M_{\mathrm{ph}}(T)$.
In the weak coupling, 
 the weak-boson masses
$M_{\sub{W,Z}}(T)$ at finite temperature 
are given by simply replacing the VEV
at the vacuum ($v_0=\mu_0/\sqrt{\lambda}=246$ GeV)
with that at  finite temperature $v(T)$:
\begin{align}
&\bigl(
M_{\sub{W}}(T),~
M_{\sub{Z}}(T),~
M_{\mathrm{ph}}(T)
\bigr)\nn\\
&=
\Bigl(
\frac{gv(T)}{2},~
\frac{\sqrt{g^2 + g^{\prime 2}}~v(T)}{2},~
0
\Bigr)\ .\label{eq:MGT}
\end{align}
In fact, 
the weak-boson masses (\ref{eq:MGT}) acquire
thermal corrections of order $gT$ from the one-loop self-energy,
as was shown by C. Manuel \cite{Manuel:1998vg} 
under the condition $M_{\sub{W,Z}}(T) \ll T$ and $\lambda \to\infty$
in the Stueckelberg formalism.
However, our main focus is on the region satisfying
$M_{\sub{W,Z}}\sim gv(T)\gg gT$, or equivalently $v(T)\gg T$,
which means that the thermal corrections to 
the masses are of higher order at weak coupling,
and can be neglected in the region of our interest.

\begin{figure}[ht]
\begin{center}
\includegraphics[width=7.5cm]{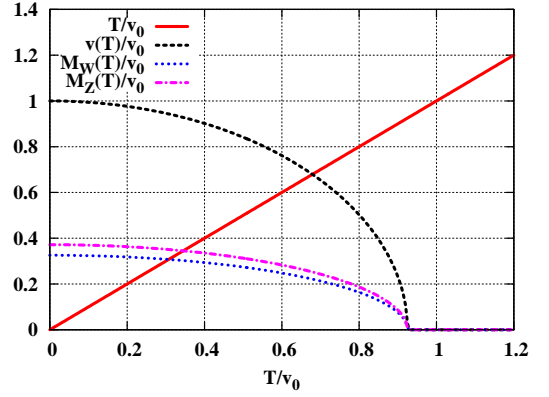}
\caption{The VEV and gauge-boson masses
in thermal background as a function of temperature $T$.
All quantities are normalized by the vacuum VEV $v_0=246$ GeV.}
\label{Fig:VEV}
\end{center}
\end{figure}

Next, we discuss how to choose 
the effective potential in the Higgs sector.
The lattice Monte Carlo simulations
have shown that 
the electroweak symmetry breaking or restoration
is a smooth crossover \cite{Lat_EWPT},
for a Higgs mass $m_{\sub{H}}$ larger than about $\sim 70$ GeV
in the standard model.
Assuming that the new boson with the mass around $125$--$127$ GeV
observed at the LHC
\cite{:2012gu,:2012gk,Aaltonen:2012if,Abazov:2012qya}
is the Higgs particle,
the possibility of
a strong first-order electroweak transition is excluded
within the standard model.
Note that a strong first-order electroweak transition
is not necessary in the thermal leptogenesis scenario, 
unlike some of other scenarios
of baryon number creation~\cite{Rubakov:1996vz}.
As pointed out in Ref.~\cite{Boyanovsky:2005hk},
the absence of the first-order transition
indicates that the weak-boson masses in electroweak plasma
$M_{\sub{W,Z}}(T)$ smoothly go to zero.
Therefore, there should exist a temperature regime
satisfying $M_{\sub{W,Z}}(T)\sim T$,
where we expect a nontrivial
spectral property of the standard model particles.
In order to implement this feature with a smooth transition,
we use the following Higgs effective potential~\cite{Kapusta}:
\begin{align}
V_{\mathrm{eff}}
&=
-\frac{1}{2}\mu^2(T)v^2(T)
+ \frac{\lambda}{4}v^4(T)
\ ,\label{eq:Veff}\\
\mu^2(T)
&=
\mu_0^2
\Bigl(
1-\frac{T^2}{T_c^2}
\Bigr)\ ,\\
T_{c}^2
&= 
\frac{4\mu_0^2}{2\lambda + 3g^2/4 + g^{\prime~2}/4}\ ,
\end{align}
where $\mu_0^2=m_{\sub{H}}^2/2$
and $\lambda=m_{\sub{H}}^2/(2v_0^2)$
are the mass and coupling
parameters in the Higgs Lagrangian (\ref{eq:LH}).
This effective potential has been derived in the $R_{\xi}$ gauge
by taking account of
the leading order of the high-$T$ expansion
for thermal one-loop effects of the Higgs and gauge bosons
in addition to the tree-level Higgs potential.
The $\xi$ dependencies cancel out
between the Nambu-Goldstone bosons and the ghost contributions.
The effective potential leads to the second-order phase transition,
and the VEV at finite $T$ is obtained as
\begin{align}
v^2(T) = v_0^2
\Bigl(1-\frac{T^2}{T_c^2}\Bigr)\ ,\quad
T\leq T_c
\ .\label{eq:VEV}
\end{align}
Figure \ref{Fig:VEV} shows
the $v(T)$ and the weak-boson masses as a function of temperature.

\subsection{Self-energy and spectral density
for left-handed neutrinos at finite $T$}
\label{subsec:SEne_lepton}
In the broken phase of 
the electroweak symmetry,
the Lagrangian for the lepton sector takes the following form:
\begin{align}
\mathcal{L}_{\sub{L}}
&=
\sum_{i=e,\mu,\tau}\Bigl[
(\bar{\nu}^i,\bar{l}_{\sub{L}}^i)
i\sla{-1.0}{-0.5}{\partial}
\begin{pmatrix}
\nu^i \\
l_{\sub{L}}^i
\end{pmatrix}
+ \bar{l}_{\sub{R}}^i
i\sla{-1.0}{-0.5}{\partial}
l_{\sub{R}}^i
\Bigr]\nn\\
&\quad +
\Bigl(
W^{\dagger}_{\mu}J^{\mu}_{\sub{W}}
+J^{\dagger}_{\sub{W}~\mu}W^{\mu}
+Z_{\mu}J^{\mu}_{\sub{Z}}
+A_{\mu}^{\sub{EM}}J^{\mu}_{\sub{EM}}
\Bigr)\ ,\label{eq:L_Lep}
\end{align}
where
$\nu^i=\mathcal{P}_{\sub{L}}\nu^i_{\sub{D}}$
and
$l_{\sub{L/R}}^i=\mathcal{P}_{\sub{L/R}}l^i$
represent the left-handed neutrinos
and left-handed (right-handed) charged leptons,
respectively.
Since we are interested in the temperature region comparable
to the weak-boson masses $T\sim M_{\sub{W,Z}}(T)$,
where the lepton masses 
are much smaller than $M_{\sub{W,Z}}(T)$,
we have neglected such  tiny masses of the leptons.
It is expected that the massless approximation will not
affect the basic spectral properties of the neutrinos, as is 
demonstrated in the case of light 
quarks in QGP~\cite{Kitazawa:2007ep,Hidaka:2011rz}.

\begin{figure*}[ht]
\begin{center}
\includegraphics[width=15.0cm]{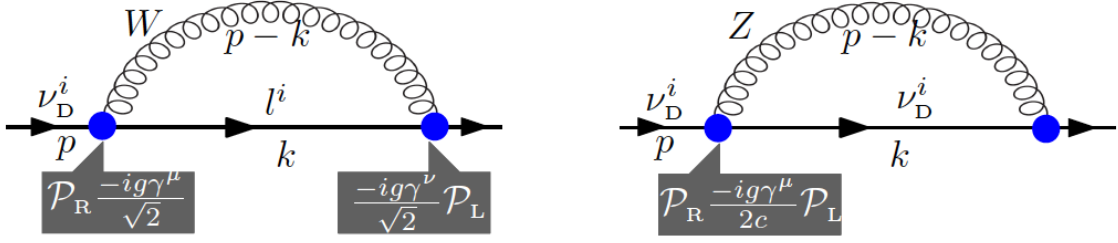}
\caption{
One-loop Feynman diagrams
of the neutrino self-energies with the weak-boson exchanges.
The symbols ``$l^i$'' and ``$\nu_{\sub{D}}^i$''
represent a charged lepton and
a Dirac neutrino in the ``$i$th'' generation, respectively.
The oriented lines represent a fermion propagator.
The chirality is managed by the projection
operators $\mathcal{P}_{\sub L/R}=(1\mp \gamma_5)/2$
on the vertex.
In the right panel, $c=\cos\theta_w$.
}\label{Fig:FD_SEneNeu}
\end{center}
\end{figure*}

The gauge currents in the Lagrangian (\ref{eq:L_Lep}) are found to be
\begin{align}
J_{\sub{W}}^{\mu}
&=
\frac{g}{\sqrt{2}}\sum_{i=e,\mu,\tau}
\bar{l}^i\gamma^{\mu}\mathcal{P}_{\sub{L}}
\nu^i_{\sub{D}}
\ ,\label{eq:JW}\\
J_{\sub{Z}}^{\mu}
&=
\sum_{i=e,\mu,\tau}\Bigl[
\frac{g\sin^2\theta_{w}}{\cos\theta_{w}}
\bar{l}^i\gamma^{\mu}l^i
\nn\\
&\quad +
\frac{g}{\cos\theta_{w}}
(\bar{\nu}^i_{\sub{D}},\bar{l}^i)
\mathcal{P}_{\sub{R}}\gamma^\mu\mathcal{P}_{\sub{L}}
\frac{\sigma^3}{2}
\begin{pmatrix}
\nu^i_{\sub{D}}\\
l^i
\end{pmatrix}
\Bigr]
\ ,\label{eq:JZ}\\
J_{\sub{EM}}^{\mu}
&=
-e\sum_{i=e,\mu,\tau}\bar{l}^i\gamma^{\mu}l^i
\ ,\label{eq:JEM}
\end{align}
where the couplings $g$, $g^{\prime}$, $e$
and the Weinberg angle $\theta_w$
are given in Table \ref{tab:params} in Appendix \ref{app:EW_params}.
The Feynman diagram for the neutrino self-energy
is shown in Fig.~\ref{Fig:FD_SEneNeu}.
We note that a neutral-current tadpole diagram with
a lepton loop does not appear
in the $CP$-symmetric thermal background, which we assume
because we are motivated by the scenario of thermal leptogenesis,
where the $CP$ violation is considered to arise from
the decay of right-handed neutrinos
rather than the thermal background itself.
For the massless neutrinos,
it is sufficient to consider the single-generation case,
and thus we drop the generation index ``$i$.''

The Feynman diagrams listed in Fig.~\ref{Fig:FD_SEneNeu}
include the following one-loop integral as a common factor,
\begin{align}
&\sigma(\mathbf{p},i\omega_m;T,M(T))\nn\\
&\quad=T\sum_n\int\frac{d^3k}{(2\pi)^3}
(-i\gamma^{\mu})
G_{\mathrm{F}}(\mathbf{k},\omega_n)
(-i\gamma^{\nu})\nn\\
&\qquad\times G_{\mu\nu}(\mathbf{p-k},i\omega_m-i\omega_n)
\ ,\label{eq:SEne_common}
\end{align}
where $G_{\mathrm{F}}$ represents an imaginary-time
propagator for a lepton.
Self-energy corrections 
in $G_{\mathrm{F}}$
are of order $g^2T$ and can be neglected  at weak coupling.
Accordingly we can use the massless propagator as $G_{\mathrm{F}}$,
\begin{align}
\label{eq:GF}
G_{\mathrm{F}}(\mathbf{k},\omega_n)
&=
\frac{i}{i\omega_n\gamma^0-\mathbf{k}\cdot\boldsymbol{\gamma}}
= \sum_{s=\pm}\frac{\Lambda_{\mathbf{k},s}\gamma^0}{i\omega_n-s\abs{k}},
\end{align}
with
\begin{align}
\Lambda_{\pm,\mathbf{k}}
&=
\frac{1\pm\gamma^0\hat{\mathbf{k}}\cdot\boldsymbol{\gamma}}{2}
\ ,\quad\quad
\hat{\mathbf{k}}=\mathbf{k}/\abs{k} \ .
\label{eq:Lambda}
\end{align}
The weak-boson propagator $G_{\mu\nu}$
is given by Eq.~(\ref{eq:G})
with the momentum identification
$q_{\mu} = (i\omega_m - i\omega_n,\mathbf{p} - \mathbf{k})$,
and the mass $M(T)$ is interpreted to be
the $W$ or $Z$ mass shown in Eq.~(\ref{eq:MGT}).
We note that
the projection operator $\mathcal{P}_{\sub{L/R}}$
has not been included in Eq.~(\ref{eq:SEne_common}),
and will be taken account of later.

The nontrivial thermal effect
emerges from the one-loop momentum integral
in the self-energy (\ref{eq:SEne_common}),
and we evaluate it
in the imaginary-time formalism.
Carrying out the summation over the Matsubara frequency $n$
and applying the analytic continuation
$i\omega_m\to \omega+i\eta$ in Eq.~(\ref{eq:SEne_common}),
we obtain the retarded self-energy as
\begin{align}
\sigma^{\mathrm{ret}}(\mathbf{p},\omega;T,M)
= \sigma^{\mathrm{ret}}_{\sub{U}}(\mathbf{p},\omega;T,M)
+ \sigma^{\mathrm{ret}}_{\xi}(\mathbf{p},\omega;T,M)
\ ,\label{eq:sig_ret}
\end{align}
where
\begin{align}
&\sigma^{\mathrm{ret}}_{\sub{U}}(\mathbf{p},\omega;T,M)\nn\\
&\quad= -\sum_{s,t = \pm}\int\frac{d^3k}{(2\pi)^3}
\frac{t\gamma^{\mu}\Lambda_{s,\mathbf{k}}\gamma^0\gamma^{\nu}}
{2E_{\mathbf{q}}}
\biggl(
g_{\mu\nu}-\frac{q_{\mu}q_{\nu}}{M^2(T)}
\biggr)\nn\\
&\qquad \times
\frac{N_{\mathrm{F}}(s\abs{k}/T) + N_{\mathrm{B}}(-tE_{\mathbf{q}}/T)}
{\omega - s\abs{k} -tE_{\mathbf{q}}(M) + i\eta}
\ ,\label{eq:sig_ret_U}\\
&\sigma^{\mathrm{ret}}_{\xi}(\mathbf{p},\omega;T,M)\nn\\
&\quad= -\sum_{s,t = \pm}\int\frac{d^3k}{(2\pi)^3}
\frac{t\gamma^{\mu}\Lambda_{s,\mathbf{k}}\gamma^0\gamma^{\nu}}
{2E_{\mathbf{q}}^{\xi}(M)}\frac{q_{\mu}q_{\nu}}{M^2(T)}\nn\\
&\qquad \times
\frac{N_{\mathrm{F}}(s\abs{k}/T) + N_{\mathrm{B}}(-tE_{\mathbf{q}}^{\xi}/T)}
{\omega - s\abs{k} -t E_{\mathbf{q}}^{\xi}(M) + i\eta}
\ .\label{eq:sig_ret_xi}
\end{align}
Here, the indices ``$U$'' and ``$\xi$"
are associated with the unitary gauge and
the $\xi$-dependent gauge-boson propagators
defined in Eqs.~(\ref{eq:G})--(\ref{eq:Gxi}).
The gauge-boson energy
$E_{\mathbf{q}}$ is defined in Eq.~(\ref{eq:E}), 
and $N_{\mathrm{F,B}}$ represent
the Fermi-Dirac and Bose-Einstein distribution functions, respectively,
\begin{align}
N_{\mathrm{F}}(x)
= \frac{1}{e^x + 1}
\ ,\quad
N_{\mathrm{B}}(x)
= \frac{1}{e^x - 1}
\ .\label{eq:Dist}
\end{align}

Similarly to the propagator $G_{\mathrm{F}}$,
we decompose
the retarded self-energy $\sigma^{\mathrm{ret}}$ as
\begin{align}
\sigma^{\mathrm{ret}}(\mathbf{p},\omega;T,M(T))
= \sum_{s=\pm}\bigl[\Lambda_{s,\mathbf{p}}\gamma^0\bigr]\
\sigma_s(\mathbf{p},\omega;T,M(T))
\ ,\label{eq:chi_decomp}
\end{align}
where the coefficients
\begin{align}
\sigma_{\pm}(\mathbf{p},\omega;T,M(T))
=
\frac{1}{2}\mathrm{tr}_{\mathrm{spin}}
\Bigl[
\Sigma^{\mathrm{ret}}(\mathbf{p},\omega;T)
\Lambda_{\pm,\mathbf{p}}\gamma^0
\Bigr]
 \label{eq:sig_pm}
\end{align}
are responsible for finite-$T$ effects.
By using $\sigma_{\pm}$,
the retarded self-energy for
the left-handed neutrinos
$\Sigma_{\mathrm{ret}}^{(\nu)}$
is now constructed as
\begin{align}
\Sigma_{\mathrm{ret}}^{(\nu)}(\mathbf{p},\omega;T)
&=
\sum_{s=\pm}
\bigl[\mathcal{P}_{\sub{R}}
\Lambda_{s,\mathbf{p}}\gamma^0
\mathcal{P}_{\sub{L}}\bigr]\
\Sigma_{s}^{(\nu)}(\abs{p},\omega;T)
\ ,\label{eq:Sig_nu_ret}
\end{align}
where
\begin{align}
&\Sigma_{\pm}^{(\nu)}(\abs{p},\omega;T)
=\biggl(\frac{g}{\sqrt{2}}\biggr)^2
\sigma_{\pm}(\abs{p},\omega;T,M_{\sub{W}}(T))\nn\\
&\quad + 
\biggl(\frac{g}{2\cos\theta_{w}}\biggr)^2
\sigma_{\pm}(\abs{p},\omega;T,M_{\sub{Z}}(T))
\ ,\label{eq:sig_nu_pm}
\end{align}
as read off from
the structure of the chirality and the gauge interaction given
by Eqs.~(\ref{eq:L_Lep})--(\ref{eq:JEM})
(and Fig.~\ref{Fig:FD_SEneNeu}).
In Eq.~(\ref{eq:Sig_nu_ret}),
the left-handed nature of massless neutrinos
is taken care of by the spinor matrix
$
\mathcal{P}_{\sub{L}}
\Lambda_{\pm,\mathbf{p}}\gamma^0
\mathcal{P}_{\sub{R}}\
$.

We shall now investigate
the neutrino spectral density,
which is defined by the imaginary part
of the retarded Green function,
\begin{align}
\rho^{(\nu)}(\mathbf{p},\omega;T)
&=
\sum_{s=\pm}
\bigl[\mathcal{P}_{\sub{R}}
\Lambda_{s,\mathbf{p}}\gamma^0
\mathcal{P}_{\sub{L}}\bigr]\
\rho_{s}^{(\nu)}(\abs{p},\omega;T)\nn\\
&\equiv
-\frac{1}{\pi}\mathrm{Im}\
G^{(\nu)}_{\mathrm{ret}}(\mathbf{p},\omega;T)
\ ,\label{eq:rho_def}
\end{align}
where
\begin{align}
G^{(\nu)}_{\mathrm{ret}}(\mathbf{p},\omega;T)
&=
\sum_{s=\pm}
\frac{\mathcal{P}_{\sub{R}}
\Lambda_{s,\mathbf{p}}\gamma^0
\mathcal{P}_{\sub{L}}}
{(\omega+i\eta)-s\abs{p}
-\Sigma^{(\nu)}_{s}(\abs{p},\omega;T)}
\ ,\label{eq:G_nu_ret}
\end{align}
or equivalently,
\begin{align}
&\rho_{\pm}^{(\nu)}(\abs{p},\omega;T)\nn\\
&= \frac{-\mathrm{Im}~\Sigma_{\pm}^{(\nu)}(\abs{p},\omega;T,\xi)/\pi
}{\{\mathcal{D}_{\pm}^{(\nu)}(\abs{p},\omega;T,\xi)\}^2
+\{\mathrm{Im}~\Sigma_{\pm}^{(\nu)}(\abs{p},\omega;T,\xi)\}^2}
\ ,\label{eq:rho_nu_pm}\\
&\mathcal{D}_{\pm}^{(\nu)}(\abs{p},\omega;T,\xi)
= \omega - \abs{p}
\mp \mathrm{Re}~\Sigma_{\pm}^{(\nu)}(\abs{p},\omega;T,\xi)
\ .\label{eq:Disp_p_nu}
\end{align}
The calculational procedure of
$\mathrm{Im}~\Sigma^{(\nu)}_{\pm}$
and
$\mathrm{Re}~\Sigma^{(\nu)}_{\pm}$
is summarized in Appendix \ref{App:SEne_sum}.

Note that finite-$T$ effects
on the spectral density $\rho^{(\nu)}$ 
of the left-handed neutrino are solely encoded in
the spectral densities $\rho_{\pm}^{(\nu)}(\abs{p},\omega;T)$
appearing as coefficients
in the decomposition (\ref{eq:rho_def}).
Thus, our focus will be put on
the spectral densities
$\rho_{\pm}^{(\nu)}(\abs{p},\omega;T)$
defined in Eq.~(\ref{eq:rho_nu_pm}).
Similarly, the thermal effects for
the spectral density of the massless charged lepton
are calculable without the complication of
the Dirac spinor structure.

Equations~(\ref{eq:sig_nu_pm}) and (\ref{eq:rho_nu_pm})
show that the computation of the neutrino spectral density
$\rho_{\pm}^{(\nu)}(\abs{p},\omega;T)$
is reduced to that of the self-energy $\sigma_{\pm}$
defined in Eq.~(\ref{eq:sig_pm}).
Fortunately,
this task is essentially the same as that of the 
self-energy of the quark coupled with massive bosonic excitations
in a QGP, and this been carried out in a series of papers
\cite{3peakQGP,Kitazawa:2007ep,Satow:2010ia}.
Therefore, we here omit the technical details
but recapitulate them
with adequate modifications for the present context in Appendices
\ref{App:Landau_Damp} and \ref{App:SEne_sum}.

\section{Neutrino spectral density:
Numerical results}\label{sec:Res}
In this section, we show our main results
for the self-energy and the spectral density
of the massless left-handed neutrinos.
In Sec. \ref{subsec:Neu_SDens_pT},
we present $\rho_+^{(\nu)}$ defined in Eq.~(\ref{eq:rho_nu_pm})
in the $\omega$-$\abs{p}$ plane at various temperatures
in {}'t Hooft-Feynman gauge ($\xi=1$), and show that 
the neutrino spectral density
has a three-peak structure in the low-energy and
low-momentum region when
the gauge-boson mass $M_{\sub{W,Z}}$ is comparable to $T$.
Next, in Sec.~\ref{subsec:Neu_SDens_3peak},
we analyze the three-peak structure in detail and the novel
low-lying collective mode for $\xi = 1$.
Finally, we examine the $\xi$ dependence
of the collective modes in Sec.~\ref{subsec:xi_dep}.
We clarify the relation between
the present results and those obtained in the previous work
\cite{Boyanovsky:2005hk}, where
the HTL approximation {is used} and the unitary gauge {is} adopted. 
All results have been obtained by using the values
in Table \ref{tab:params} in Appendix \ref{app:EW_params}.

\subsection{Neutrino spectral density
in the $\omega$-$\abs{p}$ plane
}\label{subsec:Neu_SDens_pT}

\begin{figure*}[ht]
\begin{center}
\includegraphics[width=5.8cm]{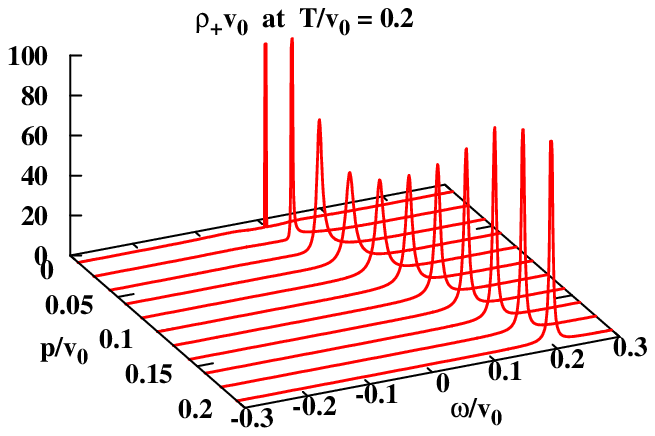}
\includegraphics[width=5.8cm]{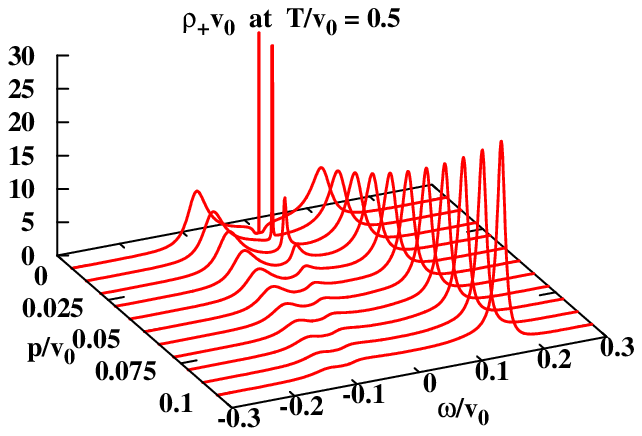}
\includegraphics[width=5.8cm]{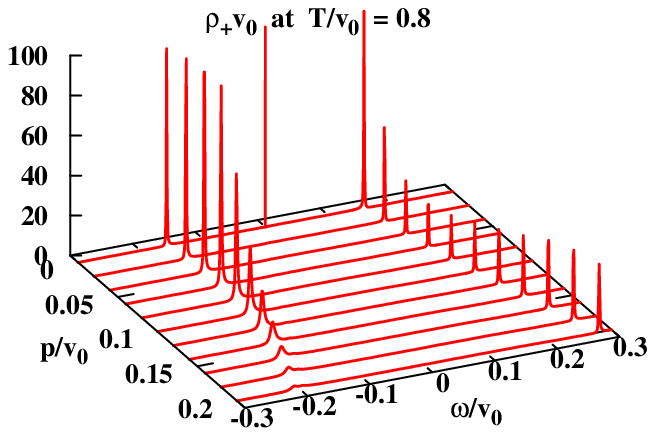}
\caption{The momentum evolution of 
the spectral density $\rho_{+}^{(\nu)}v_0$
at $T/v_0=0.2$ (left), $0.5$ (middle),
and $0.8$ (right).
}\label{Fig:SDens_SM221_1}
\end{center}
\end{figure*}

In Fig.~\ref{Fig:SDens_SM221_1},
we show the neutrino spectral density
$\rho_{+}^{(\nu)}(\abs{p},\omega;T)$
in the $\omega$-$\abs{p}$ plane
at temperatures $T/v_{0}=0.2,~0.5$, and $0.8$.
Throughout this paper,
we do not analyze $\rho_-^{(\nu)}$
because that quantity can be calculated from $\rho_+^{(\nu)}$
due to the particle-antiparticle symmetry~\cite{3peakQGP}.
In the $T/v_0=0.2$ case (left panel),
the spectral density has sharp and narrow peaks
{almost} on the light cone, $\omega = \abs{p}$.
As shown in Fig.~\ref{Fig:VEV},
the weak-boson masses at $T/v_0=0.2$ 
are still larger than the temperature,
$M_{\sub{W}}/T \simeq 1.59$,
for which the effects of thermal excitations
Eqs.~(\ref{eq:app_IDist_II}) and (\ref{eq:app_IDist_II_x})
are exponentially suppressed.
As a result, the {general} property of the spectral density
at this temperature region
is similar to that of a massless particle at zero temperature.

This feature drastically changes
when $T$ reaches around $T/v_0\sim 0.35$,
where the weak-boson masses become
comparable to $T$, as is seen in Fig.~\ref{Fig:VEV}.
At this intermediate temperature regime
($T\gtrsim M_{\sub{W,Z}}$),
the weak bosons become thermally excited with
a considerable probability.
The middle panel of Fig.~\ref{Fig:SDens_SM221_1}
shows the spectral density $\rho_+^{(\nu)}$ at $T/v_0=0.5$,
a typical example at intermediate temperature.
We see that a three-peak structure is realized 
in the low-energy and low-momentum region:
the three peaks consist of two peaks in
the positive- and negative-energy regions
and the sharp peak around the origin,
whose dispersion relations are quite different from that for the free massless
particle and are interpreted as collective excitations.
As the momentum $\abs{p}$ increases,
the peak in the positive-energy region
becomes sharper,
while the others are attenuated.
We should remark that 
a quite similar result has been obtained for the quark spectral
density in a QGP \cite{3peakQGP,Satow:2010ia},
where the quarks are
coupled with massive scalar or vector-bosonic modes in the plasma.

\begin{figure}[ht]
\begin{center}
\includegraphics[width=8.0cm]{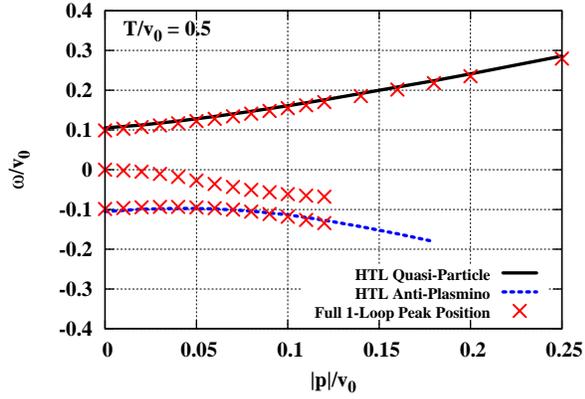}
\caption{The comparison between
the peak position of the spectral density $\rho_{+}^{(\nu)}$
shown in the middle panel of
Fig.~\protect\ref{Fig:SDens_SM221_1}
and the HTL dispersions given
by Eq.~(\protect\ref{eq:Disp_HTL}) at $T/v_0=0.5$
in the $\omega$-$\abs{p}$ plane.}
\label{Fig:SDensP_SM221_p_05}
\end{center}
\end{figure}

In order to elucidate
the physical meanings of the observed peaks at
intermediate temperature $T/v_0 = 0.5$,
we compare in Fig.~\ref{Fig:SDensP_SM221_p_05}
the peak positions of $\rho_+^{(\nu)}$
seen in the middle panel of Fig.~\ref{Fig:SDens_SM221_1}
with the massless-HTL~\cite{HTL}
dispersion relation
$\omega_{\sub{HTL}}(\abs{p})$, which is defined as
a solution of
\begin{align}
\omega_{\sub{HTL}} - \abs{p}
- \mathrm{Re}~\Sigma_{+,\sub{HTL}}^{(\nu)}
(\abs{p},\omega_{\sub{HTL}};T) = 0
\ .\label{eq:Disp_HTL}
\end{align}
Here, the real part of the self-energy
$\mathrm{Re}~\Sigma_{+,\sub{HTL}}^{(\nu)}$
is given by taking the limit
$T \gg \omega,~\abs{p},\ M_{\sub{W,Z}}$
in the full one-loop result
as detailed in Appendix \ref{App:HTL}.
Figure \ref{Fig:SDensP_SM221_p_05} shows that
the positive (negative) branch in $\rho_+^{(\nu)}$
corresponds to the quasiparticle (antiplasmino) mode, 
which is the familiar solution appearing in the massless-HTL approximation.

Although Eq.~(\ref{eq:Disp_HTL}) has a solution
in the ultrasoft region
$\omega_{\sub{HTL}}(\abs{p})\sim 0$
with $\abs{p}$ being finite,
the imaginary part in the massless-HTL approximation
$\mathrm{Im}~\Sigma_{\pm,\sub{HTL}}$ becomes huge,
and kills the peak structure at the ultrasoft region
$\omega_{\sub{HTL}}(\abs{p})\sim 0$.
Thus, the emergence of the collective modes
near the origin (ultrasoft mode) is characteristic of
the full spectral density $\rho_+^{(\nu)}$.

\begin{figure}[ht]
\begin{center}
\includegraphics[width=8.0cm]{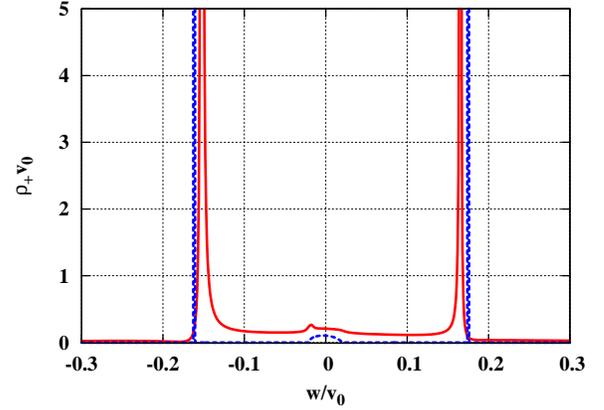}
\caption{The spectral density $\rho_{+}^{(\nu)}$ (solid red line)
as a function of $\omega$
at ($\abs{p}/v_0,T/v_0$) = ($0.02,0.8$)
compared with the spectral density obtained by using
the massless-HTL approximation (dashed blue line)
for the same conditions.}\label{Fig:SDens_SM221_w_002_08_1}
\end{center}
\end{figure}

In higher-temperature region $T\gg M_{\sub{W,Z}}$,
the two peaks in the positive- and negative-energy regions
become sharper while the central peak is greatly attenuated
[see the right panel in Fig.~\ref{Fig:SDens_SM221_1} ($T/v_0 = 0.8$)].
As shown in Fig.~\ref{Fig:SDens_SM221_w_002_08_1},
{these two} peaks approximately stay
at the same position as the quasiparticle (antiplasmino) mode
of the massless-HTL approximation.
Thus, the {general} property of the spectral density
becomes closer to that of the massless-HTL approximation
in the high-temperature region $T\gg M_{\sub{W,Z}}$.
We note that
the high-temperature region does not satisfy the condition $T\ll v(T)$;
therefore, higher-loop corrections may be non-negligible.
In fact, it is known that a resummation of 
higher-loop diagrams is necessary 
for the ultrasoft-momentum region
$\omega, \abs{p}\lesssim g^2T$~in QED and QCD \cite{Hidaka:2011rz},
where a resummation scheme is developed for obtaining  sensible results.
It would be interesting 
to develop such a resummation scheme in the electroweak theory
and investigate the spectral properties
in the vicinity of the electroweak phase transition, which is, 
however, beyond the scope of the present work and left as a future task.

\subsection{Emergence of the three-peak structure}
\label{subsec:Neu_SDens_3peak}

In this subsection, we investigate 
the properties of the three-peak structure in detail.

\begin{figure}[ht]
\begin{center}
\includegraphics[width=7.5cm]{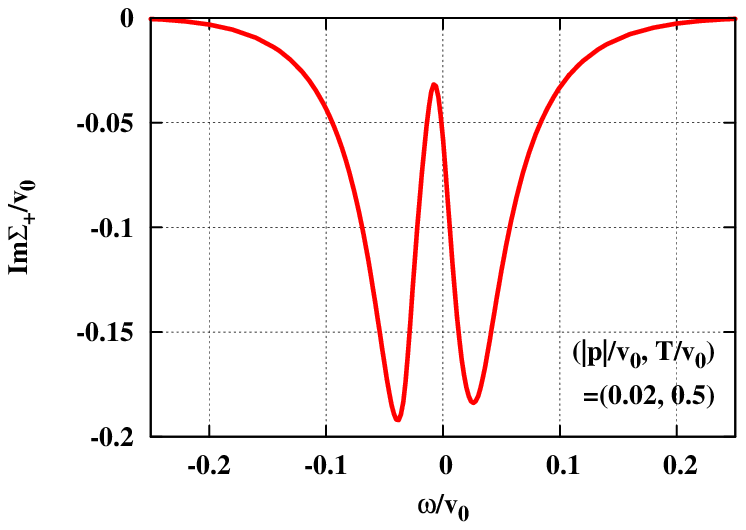}
\includegraphics[width=7.5cm]{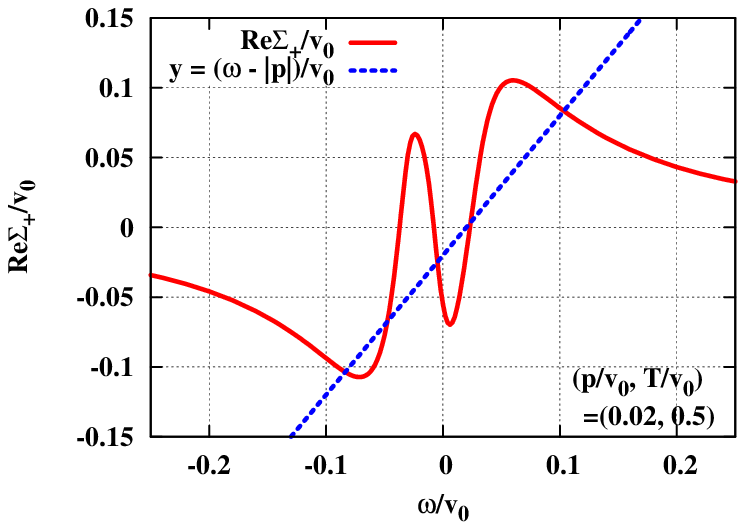}
\includegraphics[width=7.5cm]{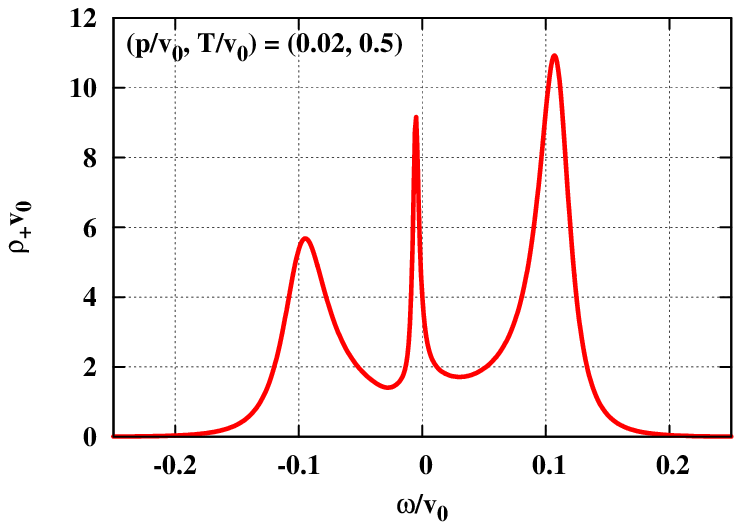}
\caption{
The self-energies
$\mathrm{Im}~\Sigma_{+}^{(\nu)}/v_0$ (upper),
$\mathrm{Re}~\Sigma_{+}^{(\nu)}/v_0$ (middle),
and the spectral density $\rho_+^{(\nu)}v_0$
at $(\abs{p}/v_0,T/v_0)=(0.02,0.5)$.}
\label{Fig:SDens_SEne_S_WZ_w_002_05_10}
\end{center}
\end{figure}

In the upper panels of
Fig.~\ref{Fig:SDens_SEne_S_WZ_w_002_05_10},
we show the imaginary part of the self-energy,
$\mathrm{Im}~\Sigma_{\pm}^{(\nu)}$
given by Eq.~(\ref{eq:Im_sig_nu_pm})
as a function of $\omega/v_0$ at
$(\abs{p}/v_0,T/v_0)=(0.02,0.5)$.
Because of the growing statistical factors
Eqs.~(\ref{eq:app_IDist_II}) and (\ref{eq:app_IDist_II_x})
with increasing temperature,
the Landau damping effect becomes significant,
and the imaginary part of the self-energy
$\mathrm{Im}~\Sigma^{(\nu)}_{+}$
drastically grows in the space-like region.
A remarkable point is that
$\mathrm{Im}~\Sigma^{(\nu)}_+$
stays relatively small in the vicinity
of $\omega\sim 0$, and
has two peaks in the space-like region
($\omega^2 \le \abs{p}^2$).
By utilizing the Kramers-Kronig-like dispersion relation (\ref{eq:Re_sig_pm_T})
(which we call the Kramers-Kronig relation to avoid the confusion
with the dispersion relation as the momentum dependence of energy),
the real part
inevitably shows an oscillatory behavior
as displayed in the middle panel
of Fig.~\ref{Fig:SDens_SEne_S_WZ_w_002_05_10}.
The crossing points with the straight dashed line
representing $y=\omega-\abs{p}$ 
correspond to the solutions of the equation,
\begin{align}
\mathcal{D}^{(\nu)}_+(\abs{p},\omega;T,\xi) = 0
\ .\label{eq:Disp_nu}
\end{align}
Here, $\mathcal{D}^{(\nu)}_+$
is defined in Eq.~(\ref{eq:Disp_p_nu}).
Two of the solutions lead to
the two peaks of the imaginary part,
and hence do not lead to
a peak in the spectral density.
The remaining three solutions contribute
to the peaks in the spectral density
(lower panel of Fig.~\ref{Fig:SDens_SEne_S_WZ_w_002_05_10}).
We note that, in the previous QGP studies \cite{3peakQGP,Satow:2010ia}, 
similar results for the real and imaginary parts of
the self-energy have been obtained, 
and the mechanism of the emergence of the three peaks in the spectral density 
was interpreted in terms of a resonant scattering~\cite{3peakQGP}.

\begin{figure}[ht]
\begin{center}
\includegraphics[width=7.5cm]{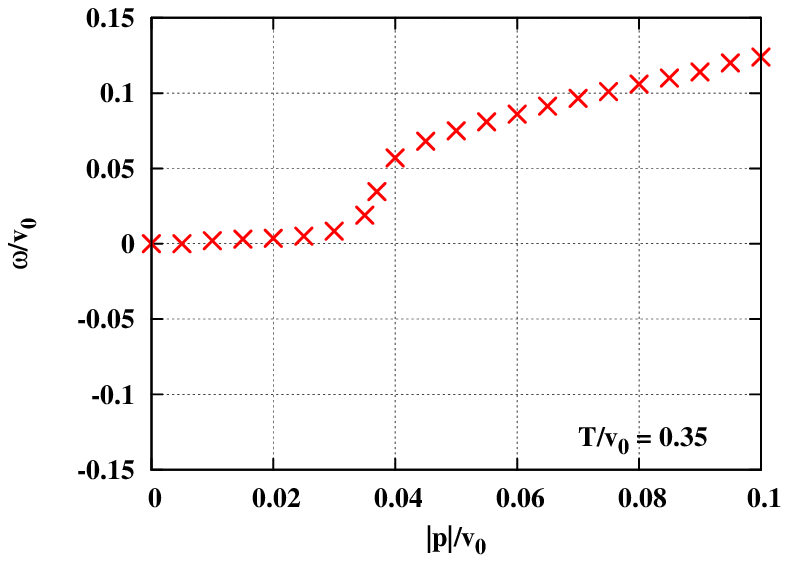}
\includegraphics[width=7.5cm]{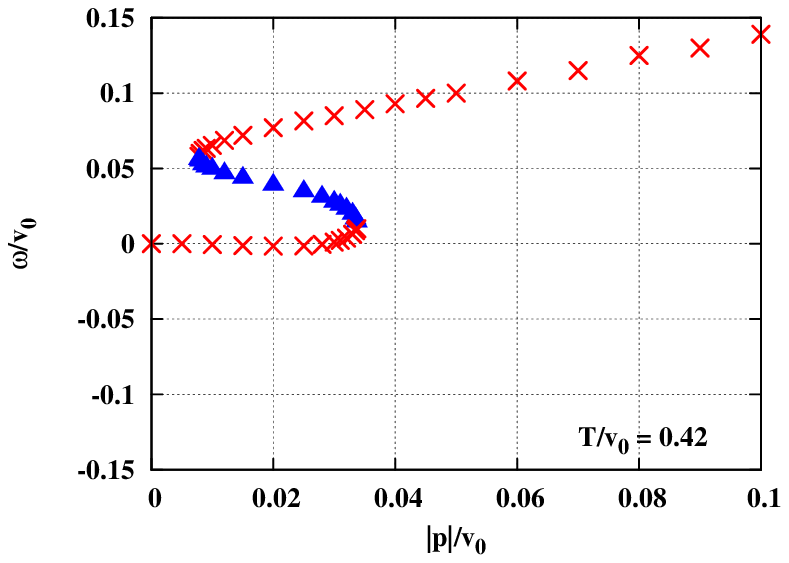}
\includegraphics[width=7.5cm]{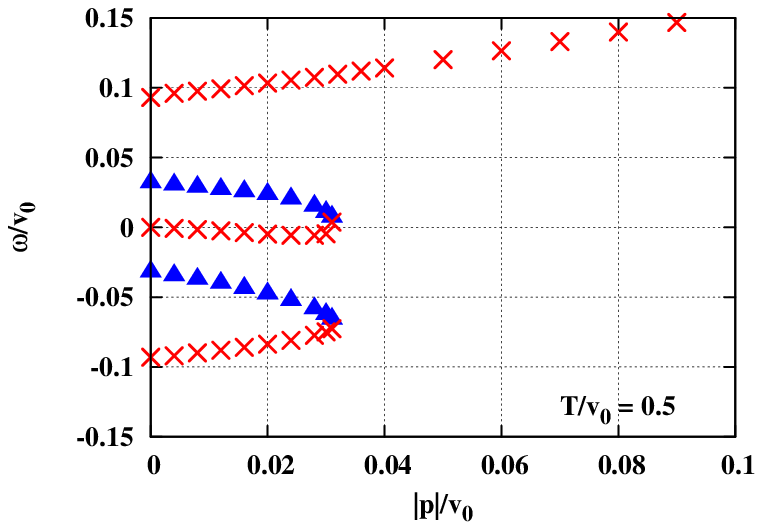}
\caption{The dispersion relations 
at $T/v_0=0.35$ (upper), 
$0.42$ (middle), and $0.5$ (lower).
The symbol $\times$
corresponds to a solution of
Eq.~(\protect\ref{eq:Disp_nu})
that gives a peak to the spectral density.
The symbol $\blacktriangle$
is a solution of
Eq.~(\protect\ref{eq:Disp_nu})
with a large imaginary part.}
\label{Fig:Disp_S_WZ_p_xx_10}
\end{center}
\end{figure}

In order to have a more complete overview,
we show the solution of Eq.~(\ref{eq:Disp_nu})
in the full $\omega$-$\abs{p}$ plane
for $T/v_0=0.35$, $0.42$, and $0.5$ cases
in Fig.~\ref{Fig:Disp_S_WZ_p_xx_10}; the resultant 
function $\omega(\abs{p})$ is called the quasidispersion relation.
The upper panel ($T/v_0=0.35$)
corresponds to the dispersion relation
at a characteristic temperature
satisfying $T/M_{\sub{W,Z}}\sim 1$,
where the Landau damping effects start to dominate.
The quasidispersion curve
deviates from the light cone, and
shows a rapid increase for
$0.03 \leq \abs{p}/v_0 \leq 0.04$.
As $T$ is increased, this steep-rise behavior is evolved to
a ``back-bending'' shape as shown in
the middle panel of Fig.~\ref{Fig:Disp_S_WZ_p_xx_10} ($T/v_0=0.42$).
Eventually, the branch with the back-bending part
crosses the $\omega$ axis
around $T/v_0\simeq 0.475$,
at which point the additional two branches appear
in the negative-$\omega$ region; see the lower panel of Fig.~\ref{Fig:Disp_S_WZ_p_xx_10}.
Thus, we get five branches in the low-momentum region.
At $\abs{p}/v_0=0.02$,
these branches correspond to the five crossing points
in the middle panel of
Fig.~\ref{Fig:SDens_SEne_S_WZ_w_002_05_10}.

We should mention that a similar branch of the
neutrino dispersion relation was found in the HTL approximation 
with the unitary gauge in Ref.~\cite{Boyanovsky:2005hk},
and the branch that touches the $\omega$ axis is
called a {\em pitchfork bifurcation}.
In Ref.~\cite{Boyanovsky:2005hk},
the pitchfork bifurcation is found to start
when the $T$-dependent mass scale
$\Delta=2M_{\sub{W}}^2(T)/(gT^2)$ exceeds
$\Delta_c\simeq 1.275$.
In the present study, the pitchfork bifurcation
starts from $T/v_0\simeq 0.475$, corresponding to
$\Delta_c\simeq 1.28$, which is consistent
with the finding in the previous
work \cite{Boyanovsky:2005hk},
and suggests the gauge independence of
the temperature at which the bifurcation starts.

\subsection{Gauge parameter $\xi$ dependence
of the three-peak structure}\label{subsec:xi_dep}
So far, we have investigated
the spectral density $\rho_{+}^{(\nu)}$
in the 't Hooft-Feynman gauge ($\xi = 1$),
and found the three-peak structure that consists of
the ultrasoft mode
as well as the quasiparticle and antiplasmino modes.
In this subsection, we investigate
the $\xi$ (in)dependence of the three-peak structure.

\begin{figure}[ht]
\begin{center}
\includegraphics[width=8.0cm]{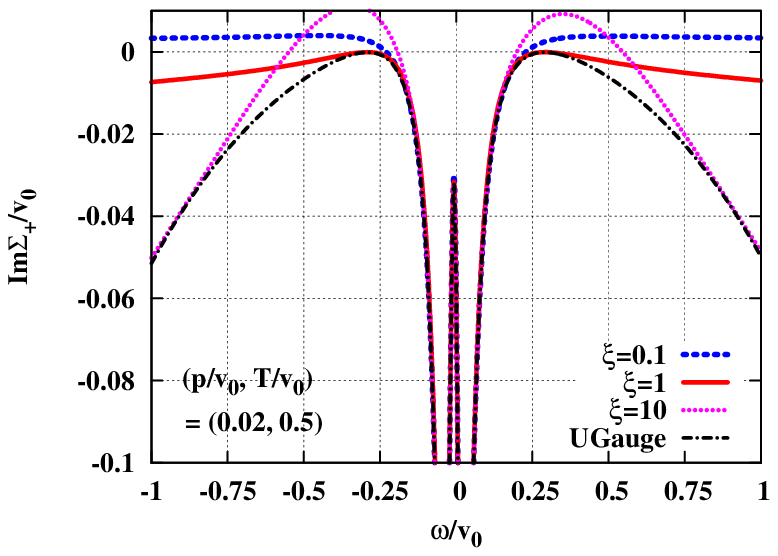}
\includegraphics[width=8.0cm]{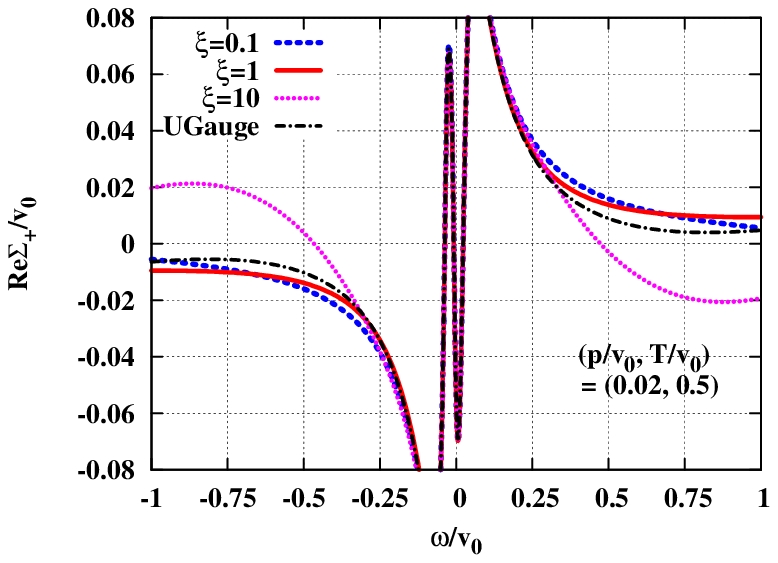}
\caption{The gauge parameter ($\xi$) dependence
of the imaginary (upper) and real (lower) parts of neutrino self-energies
at $(\abs{p}/v_0,T/v_0)=(0.02,0.5)$.
As shown in Fig.~\protect\ref{Fig:SDens_SEne_S_WZ_w_002_05_10},
the three-peak structure appears in
the low-energy region, $|\omega/v_0|\leq 0.15$,
where the $\xi$ dependence is negligible.}
\label{SEne_SM_221_w_002_05_xx}
\end{center}
\end{figure}

In Fig.~\ref{SEne_SM_221_w_002_05_xx},
we show the neutrino self-energies
for the gauge parameter $\xi = 0.1,\ 1,\ 10$.
We also display
the unitary-gauge result
which has been obtained
by utilizing the Proca formalism~\cite{3peakQGP},
which corresponds to $\xi\to \infty$.
The momentum and temperature are set to be the same as
in Fig.~\ref{Fig:SDens_SEne_S_WZ_w_002_05_10},
$(\abs{p}/v_0,T/v_0) = (0.02,0.5)$.
As shown in the figure,
the $\xi$ dependence
becomes significant with increasing $\omega$.
However, in the low-energy region $\omega/v_0\leq 0.15$
where the three-peak structure emerges
in the neutrino spectral density,
the $\xi$ dependence is found to be negligible
in both the imaginary part (upper panel)
and the real part (lower panel).
This indicates that
both the position and the width of the three peaks
are independent of the gauge parameter $\xi$.

For the three peaks of the spectral density,
the two conditions
$M_{\sub{W,Z}}\gg \omega,\abs{p}$
and
$T\gg \omega,\abs{p}$
are satisfied.
Due to the former condition,
the unitary-gauge part of the self-energy dominates
[see Eqs.~(\ref{eq:Im_sig_pm_U})--(\ref{eq:Im_sig_pm_xi})],
which accounts for the $\xi$ independence of the three-peak structure.
By using both two conditions,
we can define {an extension of the HTL approximation to the case 
where the boson mass is finite---which we call the {\em massive-HTL}
expansion---for the neutrino self-energy} as detailed in Appendix \ref{App:MHTL}.
The leading-order terms
reproduce the self-energy derived in Ref.~\cite{Boyanovsky:2005hk}.
The temperature and energy region for the pitchfork bifurcation
explained in the previous subsection is in the validity region
of this approximation scheme, which accounts for the gauge invariance
of $\Delta_c$ and the agreement of its values in the present and
the previous study \cite{Boyanovsky:2005hk}.

\section{Neutrino spectral density at $T \geq T_c$}
\label{sec:SDens_M0}
In the previous sections,
we have investigated the neutrino spectral properties
in the broken phase.
In this section, we discuss their properties
in the symmetric phase $T\geq T_c$.

In the symmetric phase,
the weak bosons become massless as is clear from Eq.(\ref{eq:MGT}).
Although higher-loop corrections for the weak-boson masses
as well as the neutrino self-energy
may become non-negligible in the symmetric phase, 
in particular, for $\omega,\abs{p}\ll g^2T$,
we dare to analyze
the neutrino self-energy within the present approximation
and  show an advantageous property of  $R_\xi$ gauge:
in the unitary gauge,
the massless limit $M\to 0$
leads to the divergence in the self-energy
due to the prefactor proportional to $1/M^2$
in the imaginary part [Eq.~(\ref{eq:Im_sig_pm_U})], and
hence the massless limit cannot be well defined
within the unitary gauge.
This problem is solved {in $R_\xi$ gauge} by 
the additional $\xi$-dependent part Eq.~(\ref{eq:Im_sig_pm_xi}):
All $\mathcal{O}(M^{-2})$ terms beautifully cancel out between
the unitary-gauge part [Eq.~(\ref{eq:Im_sig_pm_U})] and
the $\xi$-dependent part Eq.~(\ref{eq:Im_sig_pm_xi}),
and we find the finite imaginary part of the self-energy
in the vanishing gauge-boson masses
($T ,\omega, \abs{p} \gg M\to 0$) as
\begin{align}
&\mathrm{Im}~\sigma_{\pm}(\abs{p},\omega;T)|_{\sub{M}\to 0}
= \frac{\pm 1}{32\pi\abs{p}^2}\nn\\
&\quad\times\biggl[
-(1 - \xi)\abs{p}p^2
\bigl[
N_{\mathrm{F}}(X_{\mp})
+ N_{\mathrm{B}}(Y_{\mp})
\bigr]\nn\\
&\quad +
\bigl[
(1-\xi)(\omega \pm \abs{p})^2 + 2p^2
\bigr]\cdot 
TI_{\sub{ST}}(X_+,X_-;\omega/T)\nn\\
&\quad -
4(\omega \mp \abs{p})\cdot 
T^2J_{\sub{ST}}(X_+,X_-;\omega/T)
\biggr]\ ,\label{eq:Im_sig_pm_M0}
\end{align}
where $I_{\sub{ST}}$ and $J_{\sub{ST}}$ have been defined
in Eqs.~(\ref{eq:app_IDist_II}) and (\ref{eq:app_IDist_II_x}),
and their arguments $X_{\pm},Y_{\pm}$ are defined by
\begin{align}
&(X_{\pm},Y_{\pm})
\equiv\biggl(
\frac{\omega \mp \abs{p}}{2T},
\frac{-\omega \mp \abs{p}}{2T}
\biggr)\ .\label{eq:XY}
\end{align}
The technical details to derive Eq.~(\ref{eq:Im_sig_pm_M0})
are provided in Appendix \ref{App:SEne_M0}.

Furthermore,
the cancellations of $\mathcal{O}(M^{-2})$ terms
are also found
in the imaginary part of the vacuum self-energy
[$\mathrm{Im}~\sigma_{\pm,0}\equiv$
Eqs.~(\ref{eq:Im_sig_pm_U_0}) + (\ref{eq:Im_sig_pm_xi_0})],
which reads
\begin{align}
\mathrm{Im}~\sigma_{\pm,0}
(\abs{p},\omega)|_{\sub{M}\to 0}
= \frac{-\mathrm{sgn}(\omega)\xi}{16\pi}
\theta(p^2)(\omega \mp \abs{p})
\ .\label{eq:Im_sig_pm_T0M0}
\end{align}

\begin{figure}[ht]
\begin{center}
\includegraphics[width=8.0cm]{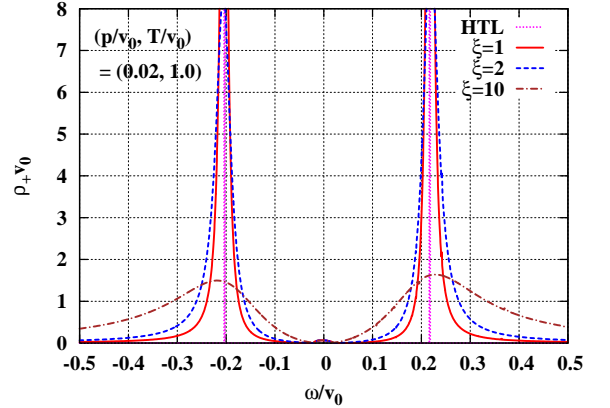}
\caption{The gauge parameter ($\xi$) dependence
of the neutrino spectral density
at $\abs{p}/v_0=0.02$
in the symmetric phase $T/v_0=1.0> T_c$.}
\label{Fig:SDens_M0221_w_002_1_xx}
\end{center}
\end{figure}

In Fig.~\ref{Fig:SDens_M0221_w_002_1_xx},
we show the neutrino spectral density at $\abs{p}/v_0=0.02$
in the symmetric phase $T/v_0=1.0> T_c$ for $\xi = 1,~2,~10$.
We find the quasiparticle and the antiplasmino peaks,
whose positions are almost independent of $\xi$
and consistent with the HTL results~\cite{quasi_particle}.
One sees that the width seems to have a  significant $\xi$ dependence.
It should be noted, however, that 
it was shown in Ref.~\cite{Satow:2010ia} that at one-loop order
the condition $\xi\lesssim 1/g$ guarantees
the smallness of the gauge dependence.
Since $g\simeq 0.7$, the upper limit of $\xi$ seems to be $\simeq 1.4$.
This estimate shows that the gauge dependence seen
in Fig.~\ref{Fig:SDens_M0221_w_002_1_xx} is consistent
with the analysis in Ref.~\cite{Satow:2010ia}
and tells us that we should restrict ourselves
to $\xi\lesssim 2$ to get a sensible result.

\section{Discussion: Toward application to resonant leptogenesis}
\label{sec:discuss}
\begin{figure*}[ht]
\begin{center}
\includegraphics[width=14.0cm]{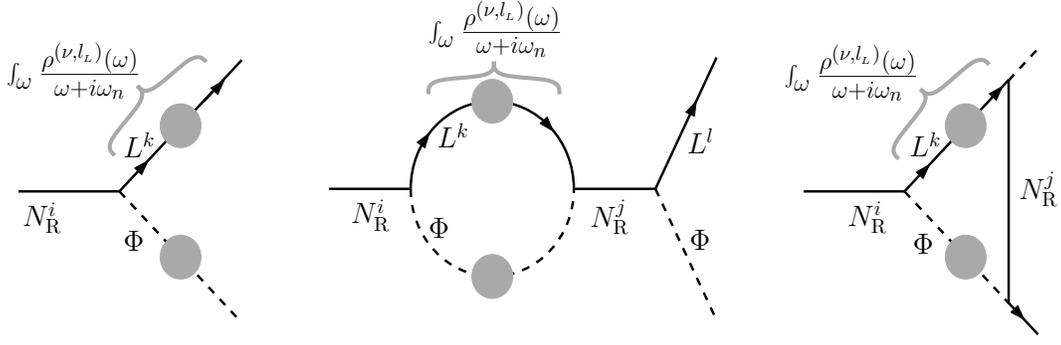}
\caption{
The diagram of the $N_{\sub{R}}^i$ decay to a 
left-handed lepton $L^i=(\nu^i,l_{\sub{L}}^i)^T$ and
a Higgs doublet $\Phi$.
Note that the lepton propagators in all the diagrams
are given in terms of the spectral densities
$\rho^{(\nu)}$ and $\rho^{(l_{\sub{L}})}$.}
\label{Fig:NR_decay_all}
\end{center}
\end{figure*}
There is a growing interest
in the collective nature of the fermion quasiparticles
in the scenario of thermal leptogenesis
\cite{Drewes:2013gca,Anisimov:2010aq,Anisimov:2010gy,Kiessig:2010pr,Besak:2012qm,Garbrecht:2013,Laine:2013vpa,Drewes:2012ma}
because the novel collective fermion modes
may significantly modify lepton number creation.
In this section,
we argue that the neutrino collective modes with
the three peaks in the spectral function obtained in the previous sections
could possibly affect the leptogenesis and hence the BAU.

In the thermal leptogenesis,
the Lagrangian is extended to include
the right-handed neutrinos ($N_{\sub{R}}^i$, $i=1,2,3$),
\begin{align}
\delta\mathcal{L}
=\bar{N}^i_{\sub{R}} i\sla{-1.0}{-0.8}\partial N_{\sub{R}}^i
+Y_{ij}\bar{N}_{\sub{R}}^i\Phi^{\dagger}L^j
-\frac{M^{ij}_{\sub{R}}}{2}\bar{N}_{\sub{R}}^i N_{\sub{R}}^j
+ \text{H.c.}\ ,\label{eq:dL}
\end{align}
where $L^i$ and $\Phi$ represent the standard model left-handed lepton
$(\nu^i,l_{\sub{L}}^i)^T$ and the standard model Higgs doublet, respectively.
The Yukawa interaction ($Y_{ij}$) term in Eq.~(\ref{eq:dL})
gives rise to  the $N_{\sub{R}}$ decay shown in Fig.~\ref{Fig:NR_decay_all},
which causes lepton number creation.
The leading effects on the finite-$T$ interaction rate
are given by the left diagram
in Fig.~\ref{Fig:NR_decay_all},
and are expressed as
\begin{align}
\Gamma_{\sub{N}}^{i}(p)
&=
-\frac{1}{2p_0}\mathrm{tr}_{\mathrm{spin}}
\Bigl[
(\sla{0.0}{-1.0}{p} + M^i_{\sub{R}})~
\mathrm{Im}\Sigma_{\sub{N}}(p)
\Bigr]\ ,\label{eq:int_rate}\\
\Sigma_{\sub{N}}(p)
&=
-4(Y^{\dagger}Y)_{11}\nn\\
&\times
T\sum_{n}\int\frac{d^3k}{(2\pi)^3}
\tilde{G}_{\mathrm{F}}(k)
\tilde{G}_{\mathrm{S}}(p-k)
\ ,\label{eq:sig_N}
\end{align}
where we have adopted the mass basis
$M^{ij}_{\sub{R}}=\delta^{ij}M^i_{\sub{R}}$,
and $\tilde{G}_{\mathrm{F}}$ and $\tilde{G}_{\mathrm{S}}$
represent the full finite-$T$ propagators of a lepton
and the Higgs particle, respectively.
The $\tilde{G}_{\mathrm{F}}$ is expressed in terms of 
the spectral density $\rho^{(\nu, l_{\sub{L}})}(\mathbf{k},\omega;T)$ 
of the left-handed neutrino $\nu$
(or the charged lepton $l_{\sub{L}}$), as follows:
\begin{align}
\tilde{G}_{\mathrm{F}}(\mathbf{k},i\omega_n;T)
=
\int_{-\infty}^{\infty}d\omega\
\frac{\rho^{(\nu, l_{\sub{L}})}(\mathbf{k},\omega;T)}
{\omega + i\omega_n}
\ .\label{eq:GF_rho}
\end{align}

At the electroweak-scale,
the spectral density $\rho^{(\nu)}$  in Eq.~(\ref{eq:GF_rho}) can have
the three-peak structure as was shown in the previous section.
Then---assuming resonant leptogenesis 
at this scale~\cite{Pilaftsis:2005rv}---these collective
modes corresponding the peaks would modify
the creation rate of the lepton number $\Gamma_{\sub{N}}^{i}$.
Moreover, the spectral density having the three peaks
is involved in both the leading and subleading diagrams
shown in Fig.~\ref{Fig:NR_decay_all},
and modifies their interference effects
leading to the $CP$ asymmetry.

Here, a natural question is whether 
the $N_{\sub{R}}$ decay into $\nu$ with the
collective nature which leads to leptogenesis
can take place before the sphaleron freeze-out.
If not, the three-peak structure would not have any relation to the BAU.
In order to study this problem,
we shall estimate the sphaleron freeze-out temperature $T_*$,
and investigate the spectral density of $\nu$ in the vicinity of $T_*$.

The net baryon number $N_{\mathrm{b}}$
is produced in the sphaleron process
when the changing rate of $N_{\mathrm{b}}$
is larger than the expanding rate of the universe $H(T)$,
\begin{align}
\Bigl|
\frac{1}{N_{\mathrm{b}}}\frac{d N_{\mathrm{b}}}{dt}
\Bigr|
\geq H(T)
\ .\label{eq:freeze_sph}
\end{align}
The freeze-out temperature $T_*$ is obtained from
the equality in Eq.~(\ref{eq:freeze_sph}).
The expanding rate $H(T)$ is given by
the Hubble parameter in the early universe at temperature $T$,
\begin{align}
H(T)
=1.66\sqrt{N_{\mathrm{dof}}}\frac{T^2}{M_{\mathrm{PL}}}
\simeq
T^2\times 1.41\times 10^{-18}\ (\mathrm{GeV})
\ ,\label{eq:Hubble}
\end{align}
with radiation degrees of freedom for the SM --
$N_{\mathrm{dof}}\simeq 106.75$,
and the Planck mass $M_{\mathrm{PL}}\simeq 1.22\times 10^{19}$ (GeV).

We evaluate the changing rate of the net baryon number \cite{Kapusta}
\begin{align}
\frac{1}{N_{\mathrm{b}}}\frac{d N_{\mathrm{b}}}{dt}
&=
-1100~
K(\lambda/g^2)~
g^7v(T)\nn\\
&\times\exp
\biggl[
-\frac{E_{\mathrm{sph}}(g,\lambda,v(T))}{T}
\biggr]\ ,\label{eq:Nb_evolv}
\end{align}
by using the same electroweak parameter set
and Higgs potential as those used
in the study of the neutrino spectral density
(see, Table \ref{tab:params}).
Under the static sphaleron background,
its free energy is known to have the expression,
\begin{align}
E_{\mathrm{sph}}=
\frac{4\pi v(T)}{g}
F_0(\lambda/g^2)\ ,\label{eq:E_sph}
\end{align}
where the dimensionless function $F_0(\lambda/g^2)$
is estimated to be
\begin{align}
F_0(\lambda/g^2)
=F_0(m_{\sub{H}}^2/(2g^2v_0^2))
\simeq F_0(0.309)\simeq 1.89\ .\label{eq:F0}
\end{align}
In the second relation --
we have used the values
listed in Table \ref{tab:params}
for $m_{\sub{H}},~v_0,~g$.
In the final relation,
we have taken
the results $F_0(0.1)=1.83$ and $F_0(1.0)=2.10$
from Table II in Ref.~\cite{Klinkhamer:1984di},
and performed a linear interpolation
to estimate $F_0(0.309)$.

The function $K(\lambda/g^2)$
is responsible for quantum fluctuations
around the sphaleron vacuum, and is well approximated
by using the expression \cite{kappa},
\begin{align}
K\Bigl(\frac{\lambda}{g^2}\Bigr)
&=
\exp\biggl[
-0.09\Bigl(\frac{\lambda}{g^2}-0.4\Bigr)^2
-0.13\Bigl(\frac{g^2}{\lambda}-2.5\Bigr)^2
\biggr]\nn\\
&\simeq 0.93\ ,\label{eq:kappa}
\end{align}
where again,
we have used the values
listed in Table \ref{tab:params}.

Substituting
Eqs.~(\ref{eq:E_sph})--(\ref{eq:kappa}) into Eq.~(\ref{eq:Nb_evolv}),
the changing rate of the net baryon number is obtained as
\begin{align}
\Big|
\frac{1}{N_{\mathrm{b}}}\frac{d N_{\mathrm{b}}}{dt}
\Big|
\simeq
51.02~v(T)~
\exp\biggl[
-36.51\frac{v(T)}{T}
\biggr]\ .\label{eq:Nb_evolv_num}
\end{align}
By using Eqs.~(\ref{eq:Nb_evolv_num}) and (\ref{eq:Hubble})
as well as Eq.~(\ref{eq:VEV}),
the inequality (\ref{eq:freeze_sph})
reduces to the following condition
\begin{align}
T\geq T_*\simeq 160~\mathrm{ GeV}
\ .\label{eq:freeze_T}
\end{align}

\begin{figure}[ht]
\begin{center}
\includegraphics[width=7.5cm]{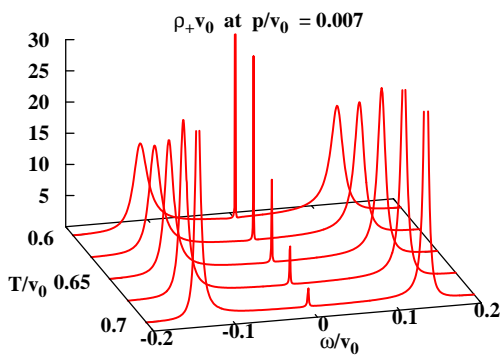}
\caption{The neutrino spectral density $\rho_{+}^{(\nu)}v_0$
in the vicinity of a sphaleron freeze-out temperature
$T_*/v_0\sim 0.65$
at a fixed low momentum $p/v_0=0.007$.
}\label{Fig:SDens_SM221_w_0007_xx_1}
\end{center}
\end{figure}

The leptons created at a temperature
satisfying $T/v_0\geq T_*/v_0\sim 0.65$ can contribute
to the BAU via the sphaleron process.
Figure~\ref{Fig:SDens_SM221_w_0007_xx_1}
shows the neutrino spectral density $\rho_{+}^{(\nu)}v_0$
for various temperatures around $T_*$
at a fixed low momentum $p/v_0=0.007$,
for which the sharp peaks are observed
in the ultrasoft region $\omega\sim 0$.
Thus, the three-peak structure is still observed in this temperature region
for small momenta, although the strength of the ultrasoft mode
may not be so strong.
This indicates that the three-peak collective modes of $\nu$
could affect leptogenesis before the freeze-out.
It is interesting that such a possibility
is shown in the definite setup.
We note that the condition $T/v_0 \sim 0.65$ 
corresponds to $T\sim v(T)$;
see Fig.~\ref{Fig:VEV}.
As explained in Sec. \ref{subsec:SEne_lepton},
higher-loop corrections
start to be non-negligible around this temperature,
and should be 
taken into account to get more sensible results.
We leave this task as a future work.

\section{Summary and concluding remarks}\label{sec:sum}
We have investigated the spectral properties 
of standard model left-handed neutrinos
without restricting {ourselves} to the HTL approximation, 
in the full energy-momentum plane
at finite $T$ around the electroweak-scale. 
This analysis is motivated by the scenario of
the electroweak-scale resonant leptogenesis
\cite{Pilaftsis,Pilaftsis:2005rv},
where the spectral property of left-handed neutrinos
can affect lepton number creation through
the decay process of right-handed neutrinos $N_{\sub{R}}$,
as shown in Fig.~\ref{Fig:NR_decay}.
We have employed the $R_{\xi}$ gauge
in order to investigate
the possible gauge-fixing dependence of the spectral density.

In the intermediate-temperature region
$M_{\sub{W,Z}}\lesssim T$
($M_{\sub{W,Z}}$ is the weak-boson mass at finite $T$),
the neutrino spectral density involves
an ultrasoft mode as well as
normal and antiplasmino modes,
and shows a three-peak structure.
The three-peak structure 
emerges independently of the gauge parameter, although a detailed
structure of the spectral density may have a slight gauge dependence.
Its emergence is accompanied
by a pitchfork bifurcation of the neutrino dispersion relation,
consistent with the previous finding in Ref.~\cite{Boyanovsky:2005hk},
where {the} HTL approximation was adopted solely in the unitary gauge.
We have mentioned that the mechanism of
the emergence of the collective excitations and the three-peak
structure is similar to what was discussed in Ref.~\cite{3peakQGP} in 
a different context.

We have examined the neutrino spectral densities
in the symmetric phase by taking
the limit of vanishing gauge-boson masses ($M_{\sub{W,Z}}\to 0$).
In this limit,
the self-energy in the unitary gauge shows a divergence,
which completely cancels out in the $R_{\xi}$ gauge.
Thus, the $R_{\xi}$ gauge fixing has allowed us to
investigate the symmetric phase.
The neutrino spectral properties in the symmetric phase
are found to be similar to that in the HTL approximation.
We have found that the possible gauge dependence of the width
of the peaks in the spectral function is
controllable, and the gauge parameter $\xi$
should be restricted to $\xi\lesssim 2$ to have a sensible result.

We have also discussed a possible implication
of the present study for particle cosmology,
in particular in the resonant leptogenesis scenario,
where it makes sense to consider a thermal leptogenesis
at electroweak-scale temperature.
We have pointed out that 
the collective modes of left-handed neutrinos
provide a novel decay channel
in the decay processes of the right-handed neutrinos,
and could modify lepton number creation.
We have estimated
the sphaleron freeze-out temperature $T_*$
and investigated 
the spectral density in the vicinity of $T_*$.
Within the present setup with
the second-order electroweak phase transition,
the ultrasoft as well as (anti)plasmino modes
can appear at a temperature comparable to $T_*$
though the strength of ultrasoft modes might not be so strong.
Thus, the novel three-peak collective modes 
could affect leptogenesis at $T\gtrsim T_*$,
and, therefore, the baryogenesis.

There are several subjects to be studied in the future:
First, it is desirable to estimate
how large the effects of two-loop or higher-order diagrams are 
on the neutrino spectral density.
Second, the present formulation should be
extended to include the bare-fermion-mass effects
\cite{Kitazawa:2007ep}.
Finally,
it would be interesting to evaluate lepton number creation
with respect to the nontrivial spectral properties
of the neutrinos as well as the charged leptons
by adopting the explicit model of resonant leptogenesis.

\section*{Acknowledgements}\label{sec:ack}
We thank Chee Sheng Fong
for fruitful discussions on resonant leptogenesis.
We thank Marco Drewes for useful comments on the quantum effects
in the thermal leptogenesis and information on some references.
This work was supported by JSPS KAKENHI Grant Numbers
20540265, 23340067, 24$\cdot$56384, and 24740184.
T.K. was partially supported
by the Yukawa International Program for Quark-Hadron Sciences,
and by a Grant-in-Aid for the global COE program
``The Next Generation of Physics,
Spun from Universality and Emergence'' from MEXT.
D. S. was supported by
JSPS Strategic Young Researcher Overseas Visits Program
for Accelerating Brain Circulation (No. R2411).
\appendix
\section{Parameters of electroweak theory}\label{app:EW_params}
We summarize the parameters of electroweak theory
and their values that were used in this paper
in Table \ref{tab:params}.
\begin{table}[hbt]
\caption{
The parameters of electroweak theory.
The entries in the third column are the values
used in this paper. The $\mathrm{SU}_{\sub{W}}(2)$
gauge coupling $g$ is chosen to reproduce
approximate experimental values of
$W$- and $Z$-boson masses at $T=0$.
}\label{tab:params}
\begin{center}
\begin{tabular}{c|c|c}
\hline\hline
Parameters & Symbols & Values in this work\\
\hline
Higgs VEV at $T=0$ &
$v_0$ &
$246$ GeV \\
Higgs mass at $T=0$ &
$m_{\sub{H}}$ &
$126$ GeV \\
Weinberg angle  &
$\theta_{w}$ &
$\sin^2\theta_{w}=0.2325$ \\
$\mathrm{SU}_{\sub{W}}(2)$ gauge coupling &
$g$ &
$0.6515$ \\
\hline
$\mathrm{U}_{\sub{Y}}(1)$ gauge coupling &
$g^{\prime}=g\tan\theta_{w}$ &
$\sim 0.3586$ \\
$\mathrm{U}_{\sub{EM}}(1)$ gauge coupling &
$e=g\sin\theta_{w}$ &
$\sim 0.3141$ \\
$W$-boson mass &
$m_{\sub{W}}=v_0g/2$ &
$\sim 80$ GeV \\
$Z$-boson mass &
$m_{\sub{Z}}=m_{\sub{W}}/\cos\theta_{w}$ &
$\sim 91$ GeV \\
\hline\hline
\end{tabular}
\end{center}
\end{table}
\section{Landau damping with massive gauge boson}
\label{App:Landau_Damp}
In this appendix,
we briefly review so-called Landau damping,
a scattering of a probe fermion by
thermally excited particles.
We closely follow Ref.~\cite{3peakQGP}.

The effects of the thermal background on
the probe particle are involved
in the one-loop retarded self-energy
$\sigma^{\mathrm{ret}}$
given by Eqs.~(\ref{eq:sig_ret})--(\ref{eq:sig_ret_xi}).
Taking the imaginary part of these equations,
we obtain
\begin{align}
&\mathrm{Im}~\sigma^{\mathrm{ret}}_{\sub{U}}(\mathbf{p},\omega;T,M)\nn\\
&= -\pi\sum_{s,t = \pm}\int\frac{d^3k}{(2\pi)^3}
\frac{t\gamma^{\mu}\Lambda_{s,\mathbf{k}}\gamma^0\gamma^{\nu}}
{2E_{\mathbf{q}}}
\biggl(
g_{\mu\nu}-\frac{q_{\mu}q_{\nu}}{M^2(T)}
\biggr)\nn\\
&\quad\times
\Bigl[N_{\mathrm{F}}(s\abs{k}/T)
+ N_{\mathrm{B}}(-tE_{\mathbf{q}}/T)\Bigr]
\delta_{\omega - s\abs{k} -tE_{\mathbf{q}}}
\ ,\label{eq:Im_sig_ret_U}\\
&\mathrm{Im}~\sigma^{\mathrm{ret}}_{\xi}(\mathbf{p},\omega;T,M)\nn\\
&= -\pi\sum_{s,t = \pm}\int\frac{d^3k}{(2\pi)^3}
\frac{t\gamma^{\mu}\Lambda_{s,\mathbf{k}}\gamma^0\gamma^{\nu}}
{2E_{\mathbf{q}}^{\xi}(M)}\frac{q_{\mu}q_{\nu}}{M^2(T)}\nn\\
&\quad\times
\Bigl[N_{\mathrm{F}}(s\abs{k}/T)
+ N_{\mathrm{B}}(-tE_{\mathbf{q}}^{\xi}/T)\Bigr]
\delta_{\omega - s\abs{k} -t E_{\mathbf{q}}^{\xi}}
\ ,\label{eq:Im_sig_ret_xi}
\end{align}
where $N_{\mathrm{F}}$ $(N_{\mathrm{B}})$ represents
the Fermi-Dirac (Bose-Einstein) distribution functions
[See Eq.~(\ref{eq:Dist})].
The delta functions provide
the momentum-conservation conditions
in the presence of massive gauge-boson effects,
\begin{align}
E_{\mathbf{q}=\mathbf{p}-\mathbf{k}}(M)&= \sqrt{\abs{q}^2+M^2(T)}
\ ,\label{eq:E_app}\\
E_{\mathbf{q}}^{\xi}(M)&\equiv E_{\mathbf{q}}(\sqrt{\xi}M)
\ .\label{eq:E_xi_app}
\end{align}
Here, $M(T)$ represents
the $W$-boson, $Z$-boson or photon mass
at finite $T$.

The physical meaning of delta functions
and distribution functions
in Eqs.~(\ref{eq:Im_sig_ret_U}) and (\ref{eq:Im_sig_ret_xi})
is more transparent when they are rewritten as
\begin{align}
&st = ++:~
\delta_{\omega - \abs{k} - E_{\mathbf{q}}}
\bigl[
(1 - N_{\mathrm{F}})(1 + N_{\mathrm{B}})
+ N_{\mathrm{F}}N_{\mathrm{B}}
\bigr]\ , \nn\\
&st = -+:~
\delta_{\omega + \abs{k} - E_{\mathbf{q}}}
\bigl[
N_{\mathrm{F}}(1 + N_{\mathrm{B}})
+ N_{\mathrm{B}}(1 - N_{\mathrm{F}})
\bigr]\ , \nn\\
&st = +-:~
\delta_{\omega - \abs{k} + E_{\mathbf{q}}}
\bigl[
N_{\mathrm{B}}(1 - N_{\mathrm{F}})
+ N_{\mathrm{F}}(1 + N_{\mathrm{B}})
\bigr]\ ,\nn\\
&st = --:~
\delta_{\omega + \abs{k} + E_{\mathbf{q}}}
\bigl[
N_{\mathrm{F}}N_{\mathrm{B}} +
(1 - N_{\mathrm{F}})(1 + N_{\mathrm{B}})
\bigr]\ ,\label{eq:cutting_T}
\end{align}
where $N_{\mathrm{F}}=N_{\mathrm{F}}(+\abs{k}/T)$ and 
$N_{\mathrm{B}}=N_{\mathrm{B}}(+E_{\mathbf{q}}/T)$
or $N_{\mathrm{B}}(+E_{\mathbf{q}}^{\xi}/T)$.
These terms can be
interpreted in terms of a scattering process of the probe
fermion with the thermal background.
A thermally excited fermion has a statistical factor
$N_{\mathrm{F}}$ in the initial state and
$(1 - N_{\mathrm{F}})$ in the final state due to Pauli blocking.
Similarly, a thermally excited boson has
a factor $N_{\mathrm{B}}$ in the initial state 
and $(1 + N_{\mathrm{B}})$ in the final state representing
the induced emission.
These thermal particles are
all on-shell in the present approximation.
We note that the second and third lines
in Eq.~(\ref{eq:cutting_T}) 
involve the thermal particles
as incident particles. 
Then the first term
in the second line of Eq.~(\ref{eq:cutting_T})
is interpreted as
a scattering process of an external lepton with
a thermally excited lepton
carrying a factor $N_{\mathrm{F}}$
into a thermally excited gauge boson
carrying a factor $1+N_{\mathrm{B}}$.
The second term is its inverse process.
The third line in Eq.~(\ref{eq:cutting_T})
describes processes with 
the incident and the final states
exchanged in the second line.
These processes are known as Landau damping,
and make decay channels in the space-like region.
The same interpretation holds for
the $\xi$-dependent part with a replacement
$E_{\mathbf{q}}\to E_{\mathbf{q}}^{\xi}$.
The Landau damping which is absent at vanishing temperature 
becomes significant at high temperature
where thermally excited particles are abundant.

For the space like-region terms
($st=-+$ and $+-$),
the momentum integral
in the unitary-gauge part [Eq.~(\ref{eq:Im_sig_ret_U})]
is evaluated as
\begin{align}
&\sum_{st=-+,+-}st\int\frac{d^3k}{(2\pi)^3}~
\delta_{\omega - s\abs{k} - tE_{\mathbf{q}}}\nn\\
&\quad=
\sum_{st=-+,+-}\frac{st}{4\pi^2}
\int_0^{\infty}d\abs{k}\abs{k}^2
\int_{-1}^{1}d(\cos\eta)~
\delta_{\omega - s\abs{k} -tE_{\mathbf{q}}}\nn\\
&\quad=
\frac{-T^3}{4\pi^2}
\biggl[
\int_{-\infty}^{x_+}
+ \int^{\infty}_{x_-}
\biggr]dx
=
\frac{T^3}{4\pi^2}
\biggl[
\int_{x_+}^{x_-}
- \int_{-\infty}^{\infty}
\biggr]dx
\ ,\label{eq:Landau_Integ}
\end{align}
where
\begin{align}
x_{\pm}=\frac{p^2 - M^2}{2T(\omega\pm\abs{p})}
\ ,\label{eq:app_x_pm}
\end{align}
and $\cos\eta$ denotes the angle between
external and internal fermion momenta
$\mathbf{p}$ and $\mathbf{k}$.
Repeating the same procedure in the time-like region,
only the contribution from the $\int_{x_+}^{x_-}$ term remains.

Substituting Eq.~(\ref{eq:Landau_Integ})
into Eq.~(\ref{eq:Im_sig_ret_U}),
the combination of the integrals
$\int_{-\infty}^{\infty}$ and/or $\int_{x_+}^{x_-}$ 
with the statistical factors
$N_{\mathrm{F}}(s\abs{k}/T)+N_{\mathrm{B}}(-tE_{\mathbf{q}}/T)$
gives rise to two types of integrals,
\begin{align}
&I_{\sub{ST}}(x_+,x_-;\omega/T)\nn\\
&\ \ \equiv
\biggl[
\int_{x_+}^{x_-} - ~
\theta(-p^2)\int_{-\infty}^{\infty}
\biggr]dx~
\bigl[
N_{\mathrm{F}}(x) +
N_{\mathrm{B}}(x-\omega/T)
\bigr]\ ,\label{eq:IDist}\\
&J_{\sub{ST}}(x_+,x_-;\omega/T)\nn\\
&\ \ \equiv
\biggl[
\int_{x_+}^{x_-} - ~
\theta(-p^2)\int_{-\infty}^{\infty}
\biggr]dx~x
\bigl[
N_{\mathrm{F}}(x) +
N_{\mathrm{B}}(x-\omega/T)
\bigr]\ .\label{eq:IDist_x}
\end{align}
Here, the step function $\theta(-p^2)$ with 
$p^2=\omega^2-\abs{p}^2$
picks up the space-like region.
The integral $\int_{-\infty}^{\infty}$
is analytically performed, and reads
\begin{align}
&I_{\sub{ST}}(x_+,x_-;\omega/T)\nn\\
&\quad=
\biggl[
\int_{x_+}^{x_-}dx~N_{\mathrm{F}}(x)
+ \int_{y_+}^{y_-}dy~N_{\mathrm{B}}(y)
\biggr]
+ \theta(-p^2)\frac{\omega}{T}
\ ,\label{eq:app_IDist_II}\\
&J_{\sub{ST}}(x_+,x_-;\omega/T)\nn\\
&\quad=
\biggl[
\int_{x_+}^{x_-}dx~xN_{\mathrm{F}}(x)
+ \int_{y_+}^{y_-}dy~yN_{\mathrm{B}}(y)
\biggr]\nn\\
&\qquad+
\frac{\omega}{T}
\int_{y_+}^{y_-}dy~N_{\mathrm{B}}(y)
- \theta(-p^2)
\Bigl[
\frac{\pi^2}{2} - \frac{\omega^2}{2T^2}
\Bigr]\ ,\label{eq:app_IDist_II_x}
\end{align}
where
\begin{align}
y_{\pm}=x_{\pm}-\omega/T
\ .\label{eq:app_y_pm}
\end{align}
The remaining integrals
$\int_{x_+}^{x_-}$
and
$\int_{y_+}^{y_-}$
will be numerically evaluated.
For the $\xi$-dependent part,
the counterparts of $I_{\sub{ST}}$ and $J_{\sub{ST}}$
are obtained by the replacements
\begin{align}
&x_{\pm}\to x_{\pm}^{\prime}
=
\frac{p^2-\xi M^2}{2T(\omega\pm \abs{p})}
\ ,\label{eq:app_x_pm_xi}\\
&y_{\pm}\to y_{\pm}^{\prime}
= x_{\pm}^{\prime}-\frac{\omega}{T}
\ .\label{eq:app_y_pm_xi}
\end{align}
As will be shown in the next appendix,
$I_{\sub{ST}}$ and $J_{\sub{ST}}$
give a key ingredient to the imaginary part of the self-energy
[see Eqs.~(\ref{eq:Im_sig_pm_U}) and (\ref{eq:Im_sig_pm_xi})].

\section{Self-energy $\sigma_{\pm}$ at finite $T$}
\label{App:SEne_sum}
Based on previous work~\cite{3peakQGP,Satow:2010ia},
we review the derivation of
the self-energy $\sigma_{\pm}$ defined in Eq.~(\ref{eq:sig_pm}).
We start from calculating the imaginary part of $\sigma_{\pm}$,
which is composed of the unitary-gauge part and
the $\xi$-dependent part,
\begin{align}
&\mathrm{Im}~\sigma_{\pm}(\abs{p},\omega;T,M) \nn\\
&\quad = \mathrm{Im}~\sigma_{\pm}^{\sub{U}}(\abs{p},\omega;T,M)
+ \mathrm{Im}~\sigma_{\pm}^{\xi}(\abs{p},\omega;T,M)
\ .\label{eq:Im_sig_pm}
\end{align}
Substituting the imaginary part of
the retarded self-energy
[Eqs.~(\ref{eq:Im_sig_ret_U}) and (\ref{eq:Im_sig_ret_xi})]
into Eq.~(\ref{eq:sig_pm}),
the straightforward computations lead to
\begin{align}
& \mathrm{Im}~\sigma_{\pm}^{\sub{U}}(\abs{p},\omega;T,M) =
\frac{\pm 1}{32\pi\abs{p}^2M^2} \nn\\
& \times\biggl[
(-p^2 + M^2)\bigl((\omega \mp \abs{p})^2 - 2M^2\bigr)\cdot
TI_{\sub{ST}}(x_+,x_-;\omega/T)\nn\\
& + 2(\omega \mp \abs{p})(p^2 - 2M^2)\cdot
T^2J_{\sub{ST}}(x_+,x_-;\omega/T)
\biggr]\ .\label{eq:Im_sig_pm_U}
\end{align}
and,
\begin{align}
& \mathrm{Im}~\sigma_{\pm}^{\xi}(\abs{p},\omega;T,M) =
\frac{\pm 1}{32\pi\abs{p}^2M^2} \nn\\
&\quad\times \biggl[
(p^2 - \xi M^2)(\omega \mp \abs{p})^2\cdot
TI_{\sub{ST}}(x_+^{\prime},x_-^{\prime};\omega/T)\nn\\
&\quad - 2p^2(\omega \mp \abs{p})\cdot
T^2J_{\sub{ST}}(x_+^{\prime},x_-^{\prime};\omega/T)
\biggr]\ ,\label{eq:Im_sig_pm_xi}
\end{align}
The integrals $I_{\sub{ST}}$ and $J_{\sub{ST}}$
are given by Eqs.~(\ref{eq:app_IDist_II}) and (\ref{eq:app_IDist_II_x}),
respectively.
In the zero-temperature limit,
keeping the relations $T < {M,\ \omega,\ \abs{p}}$,
we obtain the {\em vacuum} effects
\begin{align}
& \mathrm{Im}~\sigma_{\pm,0}^{\sub{U}}(\abs{p},\omega;M)\equiv
\mathrm{Im}~\sigma_{\pm}^{\sub{U}}(\abs{p},\omega;M,T\to 0) = \nn\\
&\quad \frac{-\mathrm{sgn}(\omega)(\omega \mp \abs{p})}{32\pi}
\Bigl(2+\frac{1}{z}\Bigr)(1-z)^2
\theta(1 - z)\ ,\label{eq:Im_sig_pm_U_0}\\
& \mathrm{Im}~\sigma_{\pm,0}^{\xi}(\abs{p},\omega;M)\equiv
\mathrm{Im}~\sigma_{\pm}^{\xi}(\abs{p},\omega;M,T\to 0) = \nn\\
&\quad \frac{\mathrm{sgn}(\omega)(\omega \mp \abs{p})}{32\pi}
\frac{(1-\xi z)^2}{z}
\theta(1 - \xi z)\ ,\label{eq:Im_sig_pm_xi_0}
\end{align}
where we have defined the dimensionless variable
$z=M^2/p^2$.

In order to make a renormalization at zero temperature, 
we extract the {\em thermal} effect
from the imaginary part of the self-energy,
\begin{align}
&\mathrm{Im}~\sigma_{\pm,T}(\abs{p},\omega;T,M)\nn\\
&\quad\equiv
\mathrm{Im}~\sigma_{\pm}(\abs{p},\omega;T,M)
- \mathrm{Im}~\sigma_{\pm,0}(\abs{p},\omega;M)
\ ,\label{eq:Im_sig_pm_T}
\end{align}
and we calculate the real part of the thermal 
and vacuum effects separately.
The former, $\mathrm{Re}~\sigma_{\pm,T}$, is computed by using
the Kramers-Kronig relation,
\begin{align}
\mathrm{Re}~\sigma_{\pm,T}(\abs{p},\omega;T)
= \mathcal{P}\int_{-\infty}^{\infty}\frac{d\omega^{\prime}}{\pi}
\frac{\mathrm{Im}~\sigma_{\pm,T}(\abs{p},\omega^{\prime};T)}
{\omega^{\prime} - \omega}
\ .\label{eq:Re_sig_pm_T}
\end{align}
Here ``$\mathcal{P}$'' represents
taking a Cauchy principal value in the integral.
The renormalization at $T=0$
can be manipulated 
by utilizing the (twice-) subtracted dispersion relation,
\begin{align}
&\mathrm{Re}~\sigma_{\pm,0}(\abs{p},\omega)
= C_0
+ C_1\cdot (\omega \mp \abs{p})\nn\\
&\quad +
(\omega \mp \abs{p})^2\cdot
\mathcal{P}\int_{-\infty}^{\infty}\frac{d\omega^{\prime}}{\pi}
\frac{\mathrm{Im}~\sigma_{\pm,0}^{\sub{U}+\xi}(\abs{p},\omega^{\prime})}
{(\omega^{\prime} \mp \abs{p})^2(\omega^{\prime} - \omega)}
\ .\label{eq:renorm}
\end{align}
The coefficients $C_{0,1}$ are determined
to be zero by imposing
on-shell renormalization conditions,
$\sigma_{\pm,0}(\abs{p},\omega)|_{\omega=\pm \abs{p}} = 0$
and
$\partial_{\omega}\sigma_{\pm,0}(\abs{p},\omega)|_{\omega=\pm \abs{p}}=0$.
The straightforward calculation leads to
\begin{align}
&\mathrm{Re}~\sigma_{\pm,0}(\abs{p},\omega)=
\frac{\omega \mp \abs{p}}{32 \pi^2}\nn\\
&\quad\times \biggl[
2z-(2+\xi)-\frac{(1-\xi z)^2}{z}
\log\Bigl|1-\frac{1}{\xi z}\Bigr|\nn\\
&\quad + \Bigl(2+\frac{1}{z}\Bigr)
(1-z)^2\log\Bigl|1-\frac{1}{z}\Bigr|
-\frac{1}{z}\log\xi
\biggr]\ ,\label{eq:Re_sig_pm_0}
\end{align}
with $z=M^2/p^2$.
The $\mathrm{Re}~\sigma_{\pm,0}$
is regular in the limit $z\to \infty$ for any finite $\xi$.

Once we obtain
$\mathrm{Im}~\sigma_{\pm}$
and
$\mathrm{Re}~\sigma_{\pm}
= \mathrm{Re}~\sigma_{\pm,T} + \mathrm{Re}~\sigma_{\pm,0}$
through the aforementioned procedures,
the neutrino self-energy
is now calculated by using Eq.~(\ref{eq:sig_nu_pm}),
\begin{align}
&\mathrm{Im}~\Sigma_{\pm}^{(\nu)}(\abs{p},\omega;T)
=\biggl(\frac{g}{\sqrt{2}}\biggr)^2
\mathrm{Im}~\sigma_{\pm}(\abs{p},\omega;T,M_{\sub{W}}(T))\nn\\
&\quad + 
\biggl(\frac{g}{2\cos\theta_{w}}\biggr)^2
\mathrm{Im}~\sigma_{\pm}(\abs{p},\omega;T,M_{\sub{Z}}(T))
\ ,\label{eq:Im_sig_nu_pm}\\
&\mathrm{Re}~\Sigma_{\pm}^{(\nu)}(\abs{p},\omega;T)
=\biggl(\frac{g}{\sqrt{2}}\biggr)^2
\mathrm{Re}~\sigma_{\pm}(\abs{p},\omega;T,M_{\sub{W}}(T))\nn\\
&\quad + 
\biggl(\frac{g}{2\cos\theta_{w}}\biggr)^2
\mathrm{Re}~\sigma_{\pm}(\abs{p},\omega;T,M_{\sub{Z}}(T))
\ .\label{eq:Re_sig_nu_pm}
\end{align}
Substituting these self-energies into Eq.~(\ref{eq:rho_nu_pm}),
we obtain the neutrino spectral density
$\rho_{\pm}^{(\nu)}(\abs{p},\omega;T)$.

\section{Massless-HTL approximation}
\label{App:HTL}
We consider the limit
$T \gg {\omega,\ \abs{p},\ M_{\sub{W,Z}}}$
in the imaginary part of the self-energy
Eqs.~(\ref{eq:Im_sig_pm})--(\ref{eq:Im_sig_pm_xi}).
In both unitary-gauge part
Eq.~(\ref{eq:Im_sig_pm_U}) and
$\xi$-dependent part [Eq.~(\ref{eq:Im_sig_pm_xi})]
the leading-order contributions are given by 
$\pi^2T^2$ terms in $T^2J_{\sub{ST}}$
[see Eq.~(\ref{eq:app_IDist_II_x}) for the expression of $J_{\sub{ST}}$].
The total imaginary part of the self-energy
in this limit reduces to
\begin{align}
& \mathrm{Im}~\sigma_{\pm,\sub{HTL}}(\abs{p},\omega;T)
=
\frac{\theta(-p^2)\pi T^2}{16\abs{p}^2}(\omega \mp \abs{p})
\ .\label{eq:Im_sig_pm_HTL}
\end{align}
The real part is analytically evaluated
through the Kramers-Kronig relation
Eq.~(\ref{eq:Re_sig_pm_T}), and reads
\begin{align}
& \mathrm{Re}~\sigma_{\pm,\sub{HTL}}(\abs{p},\omega;T) \nn\\
&\quad =
\frac{T^2}{16\abs{p}^2}\biggl(
2\abs{p} + (\omega \mp \abs{p})
\log\biggl|\frac{\omega - \abs{p}}{\omega + \abs{p}}\biggr|
\biggr)\ .\label{eq:Re_sig_pm_HTL}
\end{align}
Equations~(\ref{eq:Im_sig_pm_HTL})
and (\ref{eq:Re_sig_pm_HTL})
correspond to well-known HTL results
\cite{LeBellac,Kapusta,HTL},
and in this paper,
we call them as ``massless HTL'' approximations.

By using Eqs.~(\ref{eq:Im_sig_pm_HTL}) and (\ref{eq:Re_sig_pm_HTL}),
the neutrino self-energy in the massless-HTL is given by
\begin{align}
&\mathrm{Im}~\Sigma_{+,\sub{HTL}}^{(\nu)}(\abs{p},\omega;T)\nn\\
&\quad= 
\biggl[
\biggl(\frac{g}{\sqrt{2}}\biggr)^2
+\biggl(\frac{g}{2\cos\theta_{w}}\biggr)^2
\biggr]
\mathrm{Im}~\sigma_{+,\sub{HTL}}(\abs{p},\omega;T)
\ ,\label{eq:Im_sig_nu_p_HTL}\\
&\mathrm{Re}~\Sigma_{+,\sub{HTL}}^{(\nu)}(\abs{p},\omega;T)\nn\\
&\quad= 
\biggl[
\biggl(\frac{g}{\sqrt{2}}\biggr)^2
+\biggl(\frac{g}{2\cos\theta_{w}}\biggr)^2
\biggr]
\mathrm{Re}~\sigma_{+,\sub{HTL}}(\abs{p},\omega;T)
\ .\label{eq:Re_sig_nu_p_HTL}
\end{align}

\section{Massive-HTL expansion}\label{App:MHTL}
We derive the HTL approximation which implements
the massive weak-boson effects at high temperature
($M,T\gg \omega,\abs{p}$)
for the fermion self-energy.
We call this approximation the massive-HTL approximation,
which reproduces the unitary-gauge HTL invented by Boyanovsky
\cite{Boyanovsky:2005hk}.

As explained in Appendix \ref{App:Landau_Damp},
the one-loop momentum integral in the self-energy
gives rise to the thermal integral factor
$(I_{\sub{ST}},J_{\sub{ST}})$
given in Eqs.~(\ref{eq:app_IDist_II})--(\ref{eq:app_IDist_II_x}).
The condition $M,T\gg \omega,\abs{p}$
allows us to divide their arguments ($x_{\pm},y_{\pm}$) into
the dominant part $\zeta_{\pm}$
and the small perturbation part $\delta x_{\pm},\delta y_{\pm}$,
\begin{align}
& x_{\pm} = \zeta_{\pm} + \delta x_{\pm}\ ,\quad
y_{\pm} = \zeta_{\pm} + \delta y_{\pm}\ ,\\
& \zeta_{\pm} = \frac{-M^2}{2T(\omega\pm\abs{p})}\ ,\\
& |\delta x_{\pm}|
= \Big|\frac{\omega\mp\abs{p}}{2T}\Big| \ll 1\ ,
|\delta y_{\pm}|
= \Big|\frac{-\omega\mp\abs{p}}{2T}\Big| \ll 1\ ,\\
& |\zeta_{\pm}| \gg |\delta x_{\pm}|\ ,|\delta y_{\pm}|\ .
\end{align}
We expand the thermal integral factors
$(I_{\sub{ST}},J_{\sub{ST}})$
in terms of $\delta x_{\pm}$.
This massive-HTL expansion keeps
the mass effects in the dominant part $\zeta_{\pm}$
in sharp contrast to the massless-HTL approximation.

We perform the massive-HTL expansion in
the imaginary part of the self-energy
Eqs.~(\ref{eq:Im_sig_pm_U})--(\ref{eq:Im_sig_pm_xi}).
The leading contributions come from the unitary-gauge part
and are found to be
\begin{align}
&\mathrm{Im}~\sigma_{\pm,\sub{MHTL}}(\abs{p},\omega;T,M)
= \frac{\pm 1}{16\pi\abs{p}^2}\nn\\
&\quad\times\biggl[
-TM^2
\int_{\zeta_-}^{\zeta_+}d\zeta~
\bigl[
N_{\mathrm{F}}(\zeta)+
N_{\mathrm{B}}(\zeta)
\bigr]\nn\\
&\quad-
2T^2(\omega \mp \abs{p})
\int_{\zeta_-}^{\zeta_+}d\zeta~\zeta
\bigl[
N_{\mathrm{F}}(\zeta)+
N_{\mathrm{B}}(\zeta)
\bigr]\nn\\
&\quad+
\theta(-p^2)\cdot\pi^2T^2(\omega \mp \abs{p})
\biggr]\ .\label{eq:Im_sig_pm_MHTL}
\end{align}

We evaluate the leading effects of the real part
by substituting the imaginary part Eq.~(\ref{eq:Im_sig_pm_MHTL})
into the Kramers-Kronig relation
Eq.~(\ref{eq:Re_sig_pm_T}):
\begin{align}
&\mathrm{Re}~\sigma_{\pm,\sub{MHTL}}(\abs{p},\omega;T,M)\nn\\
&=
\mathcal{P}\int_{-\infty}^{\infty}\frac{d\omega^{\prime}}{\pi}~
\frac{\mathrm{Im~\sigma_{\pm,\sub{MHTL}}(\abs{p},\omega^{\prime};T,M)}}
{\omega^{\prime} - \omega}\nn\\
&=
\frac{\pm 1}{16\pi^2\abs{p}^2}
\int_{-\infty}^{\infty}d\omega^{\prime}
\sum_{s=\pm}s
\sum_{\mathrm{D=F,B}}\nn\\
&\qquad\times
\biggl[
TM^2 
\frac{I_{\mathrm{D}}[\zeta_{s}(\omega^{\prime})]}
{\omega^{\prime} - \omega}
+ 2T^2(\omega \mp \abs{p})
\frac{J_{\mathrm{D}}[\zeta_{s}(\omega^{\prime})]}
{\omega^{\prime} - \omega}
\biggr]\nn\\
&\quad\pm
\int_{-\infty}^{\infty}d\omega^{\prime}
\frac{\theta(-p^{\prime 2})}{16\pi^2\abs{p}^2}
\Bigl[
\pi^2T^2
\frac{\omega^{\prime} \mp \abs{p}}{\omega^{\prime} - \omega}
\Bigr]
\ ,\label{eq:Re_sig_pm_MHTL_I}
\end{align}
where, we have defined
\begin{align}
&I_{\mathrm{F,B}}[\zeta_{-}] - I_{\mathrm{F,B}}[\zeta_{+}]
\equiv
\int^{\zeta_{-}}_{\zeta_+}dx~N_{\mathrm{F,B}}(x)
\ ,\label{eq:IFB}\\
&J_{\mathrm{F,B}}[\zeta_-] - J_{\mathrm{F,B}}[\zeta_+]
\equiv
\int^{\zeta_-}_{\zeta_+}dx~xN_{\mathrm{F,B}}(x)
\ ,\label{eq:JFB}\\
&\zeta_{\pm} = -\frac{M^2}{2T(\omega \mp \abs{p})}
\ ,\label{eq:zeta_app}
\end{align}
with $p^{\prime 2} = \omega^{\prime 2} - \abs{p}^2$.
We have omitted 
$\omega$-independent terms,
which cancel one another owing to
the translation invariance of
$\int_{-\infty}^{\infty}d\omega^{\prime}$.
We introduce the new integration measure
$d\zeta=d(-M^2/(2T(\omega^{\prime} + s \abs{p})))$
with $s=\pm$, and evaluate
the integral of statistical factors $I_{\mathrm{D=F,B}}$
involved in Eq.~(\ref{eq:Re_sig_pm_MHTL_I}) as
\begin{align}
&\int_{-\infty}^{\infty}d\omega^{\prime}
\frac{I_{\mathrm{D=F,B}}[\zeta_{s}(\omega^{\prime})]}
{\omega^{\prime} - \omega}\nn\\
&\quad=
\int_{-\infty}^{\infty}\frac{d\zeta}{\zeta^2}~\frac{M^2}{2T}
\frac{I_{\mathrm{D}}[\zeta]}
{-s\abs{p} - M^2/(2T\zeta) - \omega}\nn\\
&\quad=
\int_{-\infty}^{\infty}d\zeta~
I_{\mathrm{D}}[\zeta]\cdot
\frac{d}{d\zeta}\log
\Bigl|
\frac{\omega}{T}
+ s\frac{\abs{p}}{T}
+ \frac{M^2}{2T^2\zeta}
\Bigr|\nn\\
&\quad=
\int_{-\infty}^{\infty}d\zeta~
\frac{d}{d\zeta}
\Bigl[
I_{\mathrm{D}}[\zeta]\cdot\log
\Bigl|
\frac{\omega}{T}
+ s\frac{\abs{p}}{T}
+ \frac{M^2}{2T^2\zeta}
\Bigr|
\Bigr]\nn\\
&\qquad -
\int_{-\infty}^{\infty}d\zeta~
N_{\mathrm{D}}(\zeta)\cdot
\log\Bigl|
\frac{\omega}{T}
+ s\frac{\abs{p}}{T}
+ \frac{M^2}{2T^2\zeta}
\Bigr|\nn\\
&\quad=
\Bigl[
I_{\mathrm{D}}[\infty] 
- I_{\mathrm{D}}[-\infty]
\Bigr]
\log\Bigl|
\frac{\omega}{T}
+ s\frac{\abs{p}}{T}
\Bigr|\nn\\
&\qquad -
\sum_{t=\pm}\int_{0}^{\infty}d\zeta~
N_{\mathrm{D}}(t\zeta)\cdot
\log\Bigl|
\frac{\omega}{T}
+ s\frac{\abs{p}}{T}
+ t\frac{M^2}{2T^2\zeta}
\Bigr| \ .\label{eq:II_Dist}
\end{align}
In the third equality,
we have taken the partial derivative
and used
$d I_{\mathrm{F,B}}(\zeta)/d\zeta = N_{\mathrm{F,B}}(\zeta)$.
Repeating the same procedure
for the integral of $J_{\mathrm{D=F,B}}$
contained in Eq.~(\ref{eq:Re_sig_pm_MHTL_I}),
we obtain
\begin{align}
&\int_{-\infty}^{\infty}d\omega^{\prime}
\frac{J_{\mathrm{D=F,B}}[\zeta_{s}(\omega^{\prime})]}
{\omega^{\prime} - \omega}\nn\\
&\quad=
\Bigl[
J_{\mathrm{D}}[\infty] 
- J_{\mathrm{D}}[-\infty]
\Bigr]
\log\Bigl|
\frac{\omega}{T}
+ s\frac{\abs{p}}{T}
\Bigr|\nn\\
&\qquad -
\sum_{t=\pm}\int_{0}^{\infty}d\zeta~
t\zeta\cdot
N_{\mathrm{D}}(t\zeta)\cdot
\log\Bigl|
\frac{\omega}{T}
+ s\frac{\abs{p}}{T}
+ t\frac{M^2}{2T^2\zeta}
\Bigr|\ .\label{eq:JJ_Dist}
\end{align}

Here, we utilize the following formulas
\begin{align}
&\sum_{\mathrm{D=F,B}}
\Bigl[
I_{\mathrm{D}}[\infty] 
- I_{\mathrm{D}}[-\infty]
\Bigr]\nn\\
&\quad = \int_{-\infty}^{\infty}dx~
\Bigl[
\frac{1}{e^x + 1} + \frac{1}{e^x - 1}
\Bigr]
= 0\ ,\\
&\sum_{\mathrm{D=F,B}}
\Bigl[
J_{\mathrm{D}}[\infty] 
- J_{\mathrm{D}}[-\infty]
\Bigr]\nn\\
&\quad= \int_{-\infty}^{\infty}dx~
\Bigl[
\frac{x}{e^x + 1} + \frac{x}{e^x - 1}
\Bigr]
= \frac{\pi^2}{2}\ ,\\
&\sum_{\mathrm{D=F,B}}N_{\mathrm{D}}(t\zeta)
= t\frac{2e^{-\zeta}}{1 - e^{-2\zeta}}
\ .
\end{align}
We then evaluate the summation $\sum_{t=\pm}$ in
Eqs.~(\ref{eq:II_Dist}) and (\ref{eq:JJ_Dist}) as
\begin{align}
&\sum_{\mathrm{D=F,B}}
\int_{-\infty}^{\infty}d\omega^{\prime}
\frac{I_{\mathrm{D=F,B}}[\zeta_{s}(\omega^{\prime})]}
{\omega^{\prime} - \omega}\nn\\
&\quad=
-\int_{0}^{\infty}d\zeta
\frac{2e^{-\zeta}}{1 - e^{-2\zeta}}
\log\Big|
\frac{\bar{\omega} + s\bar{\abs{p}} + \bar{\Delta}/\zeta}
{\bar{\omega} + s\bar{\abs{p}} - \bar{\Delta}/\zeta}
\Big|\ ,  \label{eq:II_Dist_II}\\
&\sum_{\mathrm{D=F,B}}
\int_{-\infty}^{\infty}d\omega^{\prime}
\frac{J_{\mathrm{D=F,B}}[\zeta_{s}(\omega^{\prime})]}
{\omega^{\prime} - \omega}\nn\\
&\quad=
\frac{\pi^2}{2}\log\Bigl|
\frac{\omega}{T}
+ s\frac{\abs{p}}{T}
\Bigr|\nn\\
&\qquad -
\int_{0}^{\infty}d\zeta
\frac{2\zeta e^{-\zeta}}{1 - e^{-2\zeta}}
\log\Big|
\frac{\bar{\omega} + s\bar{\abs{p}} + \bar{\Delta}/\zeta}
{\bar{\omega} + s\bar{\abs{p}} - \bar{\Delta}/\zeta}
\Big|\ , \label{eq:JJ_Dist_II}
\end{align}
where we have defined
$(\bar{\omega},\bar{\abs{p}}) = (\omega/T,\abs{p}/T)$
and $\bar{\Delta}=M^2/(2T^2)$.
Substituting Eqs.~(\ref{eq:II_Dist_II}) and (\ref{eq:JJ_Dist_II})
into Eq.~(\ref{eq:Re_sig_pm_MHTL_I}),
and performing $\sum_{s=\pm}$,
we obtain the real part of the massive-HTL
self-energy,
\begin{align}
&\mathrm{Re}~\sigma_{\pm,\sub{MHTL}}(\abs{p},\omega;T,M)\nn\\
&\quad= 
\frac{\pm T^2}{8\abs{p}}
\pm\frac{1}{8\pi^2\abs{p}^2}
\int_0^{\infty}d\zeta~
\frac{2\zeta e^{\zeta}}{1-e^{-2\zeta}}\nn\\
&\qquad\times
\biggl[
\Bigl(2T^2(\omega \mp \abs{p}) - \frac{TM^2}{\zeta}\Bigr)
L_{\mathrm{P}}(\bar{\omega},\bar{\abs{p}},\bar{\Delta}/\zeta)\nn\\
&\qquad+
\Bigl(2T^2(\omega \mp \abs{p}) + \frac{TM^2}{\zeta}\Bigr)
L_{\mathrm{M}}(\bar{\omega},\bar{\abs{p}},\bar{\Delta}/\zeta)
\biggr]
\ ,\label{eq:Re_sig_pm_MHTL}
\end{align}
where
\begin{align}
L_{\mathrm{P}}(\bar{\omega},\bar{\abs{p}},\bar{\Delta}/\zeta)
&=
\frac{1}{2}\log
\biggl|
\frac{\bar{\omega}+\bar{\abs{p}}+\bar{\Delta}/\zeta}
{\bar{\omega}+\bar{\abs{p}}-\bar{\Delta}/\zeta}
\biggr|\ ,\label{eq:LP}\\
L_{\mathrm{M}}(\bar{\omega},\bar{\abs{p}},\bar{\Delta}/\zeta)
&=
\frac{1}{2}\log
\biggl|
\frac{\bar{\omega}-\bar{\abs{p}}+\bar{\Delta}/\zeta}
{\bar{\omega}-\bar{\abs{p}}-\bar{\Delta}/\zeta}
\biggr|\ ,\label{eq:LM}
\end{align}
which is nothing but
the formulas in the unitary gauge
given in Ref.~\cite{Boyanovsky:2005hk}.

In summary,
the neutrino self-energy
in the leading order of the massive-HTL expansion is given by
\begin{align}
&\mathrm{Im}~\Sigma_{+,\sub{MHTL}}^{(\nu)}(\abs{p},\omega;T)\nn\\
&\quad= 
\biggl(\frac{g}{\sqrt{2}}\biggr)^2
\mathrm{Im}~\sigma_{+,\sub{MHTL}}(\abs{p},\omega;T,M_{\sub{W}})\nn\\
&\qquad+ 
\biggl(\frac{g}{2\cos\theta_{w}}\biggr)^2
\mathrm{Im}~\sigma_{+,\sub{MHTL}}(\abs{p},\omega;T,M_{\sub{Z}})
\ ,\label{eq:Im_sig_nu_p_MHTL}\\
&\mathrm{Re}~\Sigma_{+,\sub{MHTL}}^{(\nu)}(\abs{p},\omega;T)\nn\\
&\quad= 
\biggl(\frac{g}{\sqrt{2}}\biggr)^2
\mathrm{Re}~\sigma_{+,\sub{MHTL}}(\abs{p},\omega;T,M_{\sub{W}})\nn\\
&\qquad+ 
\biggl(\frac{g}{2\cos\theta_{w}}\biggr)^2
\mathrm{Re}~\sigma_{+,\sub{MHTL}}(\abs{p},\omega;T,M_{\sub{Z}})
\ .\label{eq:Re_sig_nu_p_MHTL}
\end{align}

\section{Self-energy $\sigma_{\pm}$ with vanishing gauge-boson mass}
\label{App:SEne_M0}
We analytically evaluate
the imaginary part of the self-energy at one-loop level
in the limit of vanishing gauge-boson masses
$\omega,~\abs{p}, T \gg M\to 0$,
and derive the expression (\ref{eq:Im_sig_pm_M0}).

The imaginary part of the self-energies
Eq.~(\ref{eq:Im_sig_pm}) can be rewritten as
\begin{align}
&\mathrm{Im}~\sigma_{\pm}(\abs{p},\omega;T,M(T))
=\frac{\pm 1}{32\pi\abs{p}^2} \nn\\
&\times\biggl[
\frac{-p^2}{M^2}(\omega\mp\abs{p})^2\ T
\Bigl[
I_{\sub{ST}}\Bigl(x_+,x_-;\frac{\omega}{T}\Bigl)
- I_{\sub{ST}}\Bigl(x_+^{\prime},x_-^{\prime};\frac{\omega}{T}\Bigl)
\Bigr]\nn\\
& +
(\omega\mp\abs{p})^2\ T
\Bigl[
I_{\sub{ST}}\Bigl(x_+,x_-;\frac{\omega}{T}\Bigl)
-\xi I_{\sub{ST}}\Bigl(x_+^{\prime},x_-^{\prime};\frac{\omega}{T}\Bigl)
\Bigr]\nn\\
& -
2(-p^2 + M^2)\ 
TI_{\sub{ST}}\Bigl(x_+,x_-;\frac{\omega}{T}\Bigl)\nn\\
& +
2\frac{p^2}{M^2}(\omega\mp\abs{p})\ T^2
\Bigl[
J_{\sub{ST}}\Bigl(x_+,x_-;\frac{\omega}{T}\Bigl)
- J_{\sub{ST}}\Bigl(x_+^{\prime},x_-^{\prime};\frac{\omega}{T}\Bigl)
\Bigr]\nn\\
& -
4(\omega\mp\abs{p})\
T^2J_{\sub{ST}}\Bigl(x_+,x_-;\frac{\omega}{T}\Bigl)
\biggr]\ ,\label{eq:Im_sig_pm_tot}
\end{align}
where the statistical integral factors
$I_{\sub{ST}}$ and $J_{\sub{ST}}$ are defined by 
Eqs.~(\ref{eq:app_IDist_II}) and (\ref{eq:app_IDist_II_x}),
respectively.
The arguments of $I_{\sub{ST}}$ and $J_{\sub{ST}}$
are defined as
\begin{align}
&(x_{\pm},x_{\pm}^{\prime})
= \biggl(
\frac{p^2-M^2}{2T(\omega\pm \abs{p})},
\frac{p^2-\xi M^2}{2T(\omega\pm \abs{p})}
\biggr)\ .\label{eq:app_x_pm_II}
\end{align}

In order to investigate
the gauge-boson mass dependence of the self-energy (\ref{eq:Im_sig_pm_tot})
in the limit $M\to 0$,
we expand the integral factor
($I_{\sub{ST}},J_{\sub{ST}}$) in terms of $M^2$ as
\begin{align}
&I_{\sub{ST}}(x_+,x_-)
= I_{\sub{ST}}(X_+,X_-)\nn\\
& \quad-
M^2\sum_{s=\pm}Z_{s}
\Bigl[
N_{\mathrm{F}}(X_s) + N_{\mathrm{B}}(Y_s)
\Bigr] + \mathcal{O}(M^4)
\ ,\label{eq:f_exp}\\
&J_{\sub{ST}}(x_+,x_-)
= J_{\sub{ST}}(X_+,X_-)\nn\\
& \quad-
M^2\sum_{s=\pm}Z_{s}
\Bigl[
X_sN_{\mathrm{F}}(X_s) + Y_sN_{\mathrm{B}}(Y_s)
+ \frac{\omega}{T}N_{\mathrm{B}}(Y_s)
\Bigr]\nn\\
& \quad + \mathcal{O}(M^4)
\ ,\label{eq:F_exp}
\end{align}
where, 
$x_{\pm}=X_{\pm}+M^2Z_{\pm}$, 
$(X_{\pm},Y_{\pm})
=(
(\omega \mp \abs{p})/2T,
(-\omega \mp \abs{p})/2T)$, and
$Z_{\pm}=1/(2T(\omega\pm\abs{p}))$.
%
The $M^2$ expansions for
the $\xi$-dependent sector
[$(I_{\sub{ST}}(x_+^{\prime},x_-^{\prime}),
J_{\sub{ST}}(x_+^{\prime},x_-^{\prime}))$]
are obtained by replacing $M^2$ with $\xi M^2$
in Eqs.~(\ref{eq:f_exp}) and (\ref{eq:F_exp}).

It turns out that the leading terms
$I_{\sub{ST}}(X_+,X_-)$ and $J_{\sub{ST}}(X_+,X_-)$
are common between the unitary-gauge and
the $\xi$-dependent part.
Owing to this property,
all terms proportional to $1/M^2$
cancel out in Eq.~(\ref{eq:Im_sig_pm_tot}) and read
\begin{align}
&\mathrm{Im}~\sigma_{\pm}(\abs{p},\omega;T)|_{M\to 0}
=
\frac{\pm 1}{32\pi\abs{p}^2}\nn\\
&\quad\times \biggl[
\Bigl(
(1-\xi)(\omega\mp\abs{p})^2 + 2p^2
\Bigr)\
TI_{\sub{ST}}\Bigl(X_+,X_-;\frac{\omega}{T}\Bigl)\nn\\
& \quad-
4(\omega\mp\abs{p})\
T^2J_{\sub{ST}}\Bigl(X_+,X_-;\frac{\omega}{T}\Bigl)\nn\\
&\quad -
Tp^2(1-\xi)(\omega\mp\abs{p})
\sum_{s=\pm}sZ_sH_{\pm}(X_s,Y_s)
\biggr]\ ,\label{eq:Im_sig_pm_M0_I}
\end{align}
where $H_{\pm}$ is defined as
\begin{align}
H_{\pm}(X_s,Y_s)
&=
(\omega\mp\abs{p})
\bigl[
N_{\mathrm{F}}(X_s)+N_{\mathrm{B}}(Y_s)
\bigr]
- 2\omega N_{\mathrm{B}}(Y_s)\nn\\
&\quad-
2T\bigl[
X_sN_{\mathrm{F}}(X_s)+Y_sN_{\mathrm{B}}(Y_s)
\bigr]\ .
\end{align}

Next, we simplify the final line of 
Eq.~(\ref{eq:Im_sig_pm_M0_I}).
To this end, we explicitly
evaluate $H_{\pm}(X_s,Y_s)$,
\begin{align}
H_{+}(X_-,Y_-)
&=
-2\abs{p}\bigl[
N_{\mathrm{F}}(X_-)+N_{\mathrm{B}}(Y_-)
\bigr]\ ,\\
H_{-}(X_+,Y_+)
&=
2\abs{p}\bigl[
N_{\mathrm{F}}(X_+)+N_{\mathrm{B}}(Y_+)
\bigr]\ ,\\
H_{+}(X_+,Y_+)
&=
H_{-}(X_-,Y_-)=0
\ ,\label{eq:H}
\end{align}
which enables us to perform the remaining summation
$\sum_s$ in the final line of Eq.~(\ref{eq:Im_sig_pm_M0_I}) as
\begin{align}
\sum_s sZ_sH_{\pm}(X_s,Y_s)
&=
-2\abs{p}Z_{\mp}
\bigl[
N_{\mathrm{F}}(X_{\mp})+N_{\mathrm{B}}(Y_{\mp})
\bigr]\nn\\
&=
-\frac{\abs{p}[N_{\mathrm{F}}(X_{\mp})+N_{\mathrm{B}}(Y_{\mp})]}
{T(\omega\mp\abs{p})}\ .
\end{align}
Substituting this expression
for Eq.~(\ref{eq:Im_sig_pm_M0_I}),
we obtain the desired expression (\ref{eq:Im_sig_pm_M0})
as the imaginary part of the self-energy
with vanishing gauge-boson masses.



\begin{thebibliography}{99}
\bibitem{LeBellac}
M. Le Bellac, {\em Thermal Field Theory}
(Cambridge University Press,
Cambridge, England, 1996).


\bibitem{Kapusta}
Joseph I. Kapusta and Charles Gale,
{\em Finite-Temperature Field Theory, Principles and Applications}
(Cambridge University Press, Cambridge, England, 2006).



\bibitem{Pilaftsis}
  A.~Pilaftsis,
Phys.\ Rev.\ D {\bf 56}, 5431 (1997);\,
Nucl.\ Phys.\ {\bf B504}, 61 (1997);
  Phys.\ Rev.\ Lett.\  {\bf 95}, 081602 (2005);
  A.~Pilaftsis and T.~E.~J.~Underwood,
  Nucl.\ Phys.\  {\bf B692}, 303 (2004);
  A.~Pilaftsis,
  Phys.\ Rev.\  D {\bf 56}, 5431 (1997).



\bibitem{Pilaftsis:2005rv}
  A.~Pilaftsis and T.~E.~J.~Underwood,
  Phys.\ Rev.\  D {\bf 72}, 113001 (2005).



\bibitem{Fukugita:1986hr}
  M.~Fukugita and T.~Yanagida,
  Phys.\ Lett.\  B {\bf 174}, 45 (1986).


\bibitem{LGene_Review}
For recent reviews, see for example --
  W.~Buchmuller, R.~D.~Peccei and T.~Yanagida,
  Ann.\ Rev.\ Nucl.\ Part.\ Sci.\  {\bf 55}, 311 (2005);
  C.~S.~Fong, M.~C.~Gonzalez-Garcia and E.~Nardi,
  Int.\ J.\ Mod.\ Phys.\  A {\bf 26}, 3491 (2011);
  S.~Davidson, E.~Nardi and Y.~Nir,
  Phys.\ Rep.\  {\bf 466}, 105 (2008);
  A.~Pilaftsis,
  J.\ Phys.\ Conf.\ Ser.\  {\bf 171}, 012017 (2009).


\bibitem{Kuzmin:1985mm}
  V.~A.~Kuzmin, V.~A.~Rubakov, and M.~E.~Shaposhnikov,
  Phys.\ Lett.\  B {\bf 155}, 36 (1985).


\bibitem{Klinkhamer:1984di}
  F.~R.~Klinkhamer and N.~S.~Manton,
  Phys.\ Rev.\  D {\bf 30}, 2212 (1984).


\bibitem{Arnold:1987mh} 
  P.~B.~Arnold and L.~D.~McLerran,
  Phys.\ Rev.\ D {\bf 36}, 581 (1987).



\bibitem{EBHGHK}
  F.~Englert and R.~Brout,
  Phys.\ Rev.\ Lett.\  {\bf 13}, 321 (1964);
  P.~W.~Higgs,
  Phys.\ Lett.\  {\bf 12}, 132 (1964);
  G.~S.~Guralnik, C.~R.~Hagen and T.~W.~B.~Kibble,
  Phys.\ Rev.\ Lett.\  {\bf 13}, 585 (1964).



\bibitem{quasi_particle}
  V.~V.~Klimov,
Yad.\ Fiz.\  {\bf 33}, 1734 (1981)
[Sov.\ J.\ Nucl.\ Phys.\  {\bf 33}, 934 (1981)];
  H.~A.~Weldon,
  Phys.\ Rev.\  D {\bf 40}, 2410 (1989);
  Phys.\ Rev.\  D {\bf 26}, 2789 (1982);
  Phys.\ Rev.\  D {\bf 26}, 1394 (1982).



\bibitem{HTL}
  R.~D.~Pisarski,
  Phys.\ Rev.\ Lett.\  {\bf 63}, 1129 (1989);
  E.~Braaten and R.~D.~Pisarski,
  Nucl.\ Phys.\  {\bf B337}, 569 (1990);
  Nucl.\ Phys.\  {\bf B339}, 310 (1990);
  J.~Frenkel and J.~C.~Taylor,
  Nucl.\ Phys.\  {\bf B334}, 199 (1990).




\bibitem{Baym:1992eu}
  G.~Baym, J.~P.~Blaizot and B.~Svetitsky,
  Phys.\ Rev.\  D {\bf 46}, 4043 (1992).


\bibitem{DOlivo:1992vm}
  J.~C.~D'Olivo, J.~F.~Nieves and M.~Torres,
  Phys.\ Rev.\  D {\bf 46}, 1172 (1992).



\bibitem{Boyanovsky:2005hk}
  D.~Boyanovsky,
  Phys.\ Rev.\  D {\bf 72}, 033004 (2005).


\bibitem{3peakQGP}
  M.~Kitazawa, T.~Kunihiro and Y.~Nemoto,
  Phys.\ Lett.\  B {\bf 633}, 269 (2006);
  M.~Kitazawa, T.~Kunihiro and Y.~Nemoto,
  Prog.\ Theor.\ Phys.\  {\bf 117}, 103 (2007).

\bibitem{Kitazawa:2007ep}
  M.~Kitazawa, T.~Kunihiro, K.~Mitsutani and Y.~Nemoto,
  Phys.\ Rev.\  D {\bf 77}, 045034 (2008).

\bibitem{Satow:2010ia}
  D.~Satow, Y.~Hidaka and T.~Kunihiro,
  Phys.\ Rev.\  D {\bf 83}, 045017 (2011).


\bibitem{Stueckel_O}
E.~C.~G.~Stueckelberg,
Helv.\ Phys.\ Acta {\bf 11}, 225 (1938);
Helv.\ Phys.\ Acta {\bf 11}, 299 (1938);
Helv.\ Phys.\ Acta {\bf 11}, 312 (1938).


\bibitem{Stueckel_R}
For a review article, see
H.~Ruegg and M.~Ruiz-Altaba,
Int.\ J.\ Mod.\ Phys. A{\bf 19}, 3265 (2004).


\bibitem{HaraNemo:2008}
  M.~Harada and Y.~Nemoto,
  Phys.\ Rev.\  D {\bf 78}, 014004 (2008);
  M.~Harada, Y.~Nemoto, and S.~Yoshimoto,
Prog.\ Theor.\ Phys.\ {\bf 119},\ 117 (2008).



\bibitem{Qin:2010pc}
  S.~x.~Qin, L.~Chang, Y.~x.~Liu and C.~D.~Roberts,
  Phys.\ Rev.\  D {\bf 84}, 014017 (2011).


\bibitem{Nakkagawa:2012}
H.~Nakkagawa, H.~Yokota and K.~Yoshida,
   Phys.\ Rev.\ D {\bf 85} (2012) 031902;
H.~Nakkagawa, H.~Yokota and K.~Yoshida,
  Phys.\ Rev.\ D {\bf 86} 096007 (2012).



\bibitem{Fujikawa:1972fe}
  K.~Fujikawa, B.~W.~Lee and A.~I.~Sanda,
  Phys.\ Rev.\  D {\bf 6}, 2923 (1972).


\bibitem{Manuel:1998vg}
  C.~Manuel,
  Phys.\ Rev.\  D {\bf 58}, 016001 (1998).


\bibitem{Lat_EWPT}
  K.~Kajantie, M.~Laine, K.~Rummukainen and M.~E.~Shaposhnikov,
  Phys.\ Rev.\ Lett.\  {\bf 77}, 2887 (1996);
F.~Csikor, Z.~Fodor, P.~Hegedus, A.~Jakovac, S.~D.~Katz, and A.~Piroth,
Phys.\ Rev.\ Lett.\ {\bf 85}, 932 (2000);
Y.~Aoki, F.~Csikor, Z.~Fodor, and A.~Ukawa,
Phys.\ Rev.\ D {\bf 60}, 013001 (1999);
F.~Csikor, Z.~Fodor, and J.~Heitger,
Nucl.\ Phys.\ B Proc.\ Suppl.\ {\bf 73}, 659 (1999);
Phys.\ Rev.\ Lett. {\bf 82}, 21 (1999).



\bibitem{:2012gu}
S.~Chatrchyan {\it et al.}  [CMS Collaboration],
Phys.\ Lett.\ B {\bf 716}, 30 (2012).


\bibitem{:2012gk} 
G.~Aad {\it et al.}  [ATLAS Collaboration],
Phys.\ Lett.\ B {\bf 716}, 1 (2012).


\bibitem{Aaltonen:2012if} 
  T.~Aaltonen {\it et al.}  [CDF Collaboration],
  Phys.\ Rev.\ Lett.\  {\bf 109}, 111802 (2012).


\bibitem{Abazov:2012qya} 
  V.~M.~Abazov {\it et al.}  [D0 Collaboration],
  Phys.\ Rev.\ Lett.\  {\bf 109}, 121802 (2012).

\bibitem{Rubakov:1996vz}
  V.~A.~Rubakov and M.~E.~Shaposhnikov,
  Usp.\ Fiz.\ Nauk {\bf 166}, 493 (1996)
  [Phys.\ Usp.\  {\bf 39}, 461 (1996)].

\bibitem{Hidaka:2011rz}
 V.~V.~Lebedev and A.~V.~Smilga,
  Ann. Phys.\  {\bf 202}, 229 (1990);
  Y.~Hidaka, D.~Satow, and T.~Kunihiro,
  Nucl.\ Phys.\  {\bf A876}, 93 (2012).




\bibitem{Drewes:2013gca} 
For a recent review, see,
M.~Drewes,
Int. J. Mod. Phys. E {\bf 22}, 1330019 (2013).


\bibitem{Anisimov:2010aq}
  A.~Anisimov, W.~Buchmuller, M.~Drewes and S.~Mendizabal,
  Phys.\ Rev.\ Lett.\  {\bf 104}, 121102 (2010);
  Ann. Phys.\  {\bf 326}, 1998 (2011).


\bibitem{Kiessig:2010pr}
  C.~P.~Kiessig, M.~Plumacher and M.~H.~Thoma,
  Phys.\ Rev.\  D {\bf 82}, 036007 (2010);
  C.~Kiessig and M.~Plumacher,
  J. Cosmol. Astropart. Phys. 07 (2012) 014;
  C.~Kiessig and M.~Plumacher,
  J. Cosmol. Astropart. Phys. 09 (2012) 012.



\bibitem{Besak:2012qm} 
  D.~Besak and D.~Bodeker,
  J. Cosmol. Astropart. Phys. 03 (2012) 029.

\bibitem{Anisimov:2010gy} 
  A.~Anisimov, D.~Besak and D.~Bodeker,
  J. Cosmol. Astropart. Phys. 03 (2011) 042.
  [arXiv:1012.3784 [hep-ph]].


\bibitem{Garbrecht:2013}
  B.~Garbrecht, F.~Glowna and M.~Herranen,
  J. High Energy Phys. 04 (2013) 099;
  B.~Garbrecht, F.~Glowna and P.~Schwaller,
  arXiv:1303.5498.


\bibitem{Laine:2013vpa} 
  M.~Laine,
  J. High Energy Phys. 05 (2013) 083.


\bibitem{Drewes:2012ma} 
  M.~Drewes and B.~Garbrecht,
  J. High Energy Phys. 03 (2013) 096.


\bibitem{kappa}
  L.~Carson, X.~Li, L.~D.~McLerran and R.~T.~Wang,
  Phys.\ Rev.\  D {\bf 42}, 2127 (1990);
  J.~Baacke and S.~Junker,
  Phys.\ Rev.\  D {\bf 50}, 4227 (1994);
  Phys.\ Rev.\  D {\bf 49}, 2055 (1994).

\end{thebibliography}
\end{document}